\documentclass[utf8]{FrontiersinHarvard} 

\usepackage{microtype,subcaption}
\usepackage[onehalfspacing]{setspace}

\usepackage[breaklinks, colorlinks, citecolor=blue, urlcolor=blue]{hyperref}
\usepackage{amsmath}
\usepackage{physics}
\usepackage{multirow}
\usepackage[dvipsnames]{xcolor}
\usepackage{booktabs}
\usepackage{xspace}

\newcommand{\msun}{$M_{\odot}$\xspace}

\newcommand{\nustar}{\textit{NuSTAR}\xspace}

\newcommand{\swift}{\textit{Swift}\xspace}

\newcommand{\xmm}{{\it XMM-Newton}\xspace}
\newcommand{\hexp}{\textit{HEX-P}\xspace}
\newcommand{\mbh}{${M_{\rm BH}}$}
\newcommand{\spinmass}{$M_{\rm BH}-a^\ast$}
\newcommand{\spin}{$a^\ast$}
\newcommand{\radeff}{$\eta(a^\ast)$}
\newcommand{\dspin}{$\sigma_{a^\ast}$}
\newcommand{\acc}{accretion-only}
\newcommand{\accmer}{accretion~+~mergers}
\newcommand{\bhmdot}{$\dot{M}$}

\def\totexp{10}

\def\keyFont{\fontsize{8}{11}\helveticabold }
\def\firstAuthorLast{Piotrowska~{et~al.}} 
\def\Authors{J.~M.~Piotrowska\,$^{1,*}$, 
J.~A.~Garc\'ia\,$^{2,1}$, 
D.~J.~Walton\,$^{3}$, 
R.~S.~Beckmann\,$^{4}$, 
D.~Stern\,$^{5}$,
D.~R.~Ballantyne\,$^{6}$,
D.~R.~Wilkins\,$^{7}$,
S.~Bianchi\,$^{8}$,
P.~G.~Boorman\,$^{1}$,
J.~Buchner\,$^{9}$,
C.-T.~Chen\,$^{10,11}$,
P.~Coppi\,$^{12}$,
T.~Dauser\,$^{13}$,
A.~C.~Fabian\,$^{4}$,
E.~Kammoun\,$^{14,15,8}$,
K.~Madsen\,$^{2}$, 
L.~Mallick\,$^{16,17,1}$,
G.~Matt\,$^{8}$,
G.~Matzeu\,$^{18}$,
E.~Nardini\,$^{15}$,
A.~Pizzetti\,$^{19}$,
S.~Puccetti\,$^{20}$,
C.~Ricci\,$^{21,22}$,
F.~Tombesi\,$^{23,24,25,26,2}$,
N.~Torres-Alb\`{a}\,$^{19}$,
K.-W.~Wong\,$^{27}$,
and the HEX-P Collaboration
}

\def\Address{\footnotesize
$^{1}$Cahill Center for Astronomy and Astrophysics, California Institute of Technology, Pasadena, CA, USA\\
$^{2}$X-ray Astrophysics Laboratory, NASA Goddard Space Flight Center, Greenbelt, MD, USA\\
$^{3}$Centre for Astrophysics Research, University of Hertfordshire, College Lane, Hatfield, UK\\
$^{4}$Institute of Astronomy, University of Cambridge, Madingley Road, Cambridge, UK\\
$^{5}$Jet Propulsion Laboratory, California Institute of Technology, Pasadena, CA, USA\\
$^{6}$Center for Relativistic Astrophysics, School of Physics, Georgia Institute of Technology, Atlanta, GA, USA\\
$^{7}$Kavli Institute for Particle Astrophysics and Cosmology, Stanford University, Stanford, CA, USA\\
$^{8}$Dipartimento di Matematica e Fisica, Universit\`{a} Roma Tre, Rome, Italy\\
$^{9}$Max-Planck-Institut f\"{u}r Extraterrestrische Physik (MPE), Garching bei M\"{u}nchen, Germany\\
$^{10}$Marshall Space Flight Center, Huntsville, AL, USA\\
$^{11}$Science and Technology Institute, Universities Space Research Association, Huntsville, AL, USA\\
$^{12}$Department of Astronomy, Yale University, New Haven, CT, USA\\
$^{13}$Dr. Karl Remeis-Observatory and Erlangen Centre for Astroparticle Physics, FAU Erlangen-Nürnberg, Bamberg, Germany\\
$^{14}$IRAP, Universit\'{e} de Toulouse, CNRS, UPS, CNES, Toulouse, France \\
$^{15}$INAF -- Osservatorio Astrofisico di Arcetri, Firenze, Italy \\
$^{16}$University of Manitoba, Department of Physics \& Astronomy, Winnipeg, Manitoba, Canada\\
$^{17}$Canadian Institute for Theoretical Astrophysics, University of Toronto, Toronto, Ontario, Canada\\
$^{18}$Quasar Science Resources SL for ESA, ESAC, Science Operations Department, Madrid, Spain\\
$^{19}$Department of Physics and Astronomy, Clemson University, Kinard Lab of Physics, Clemson, SC, USA\\
$^{20}$ASI - Agenzia Spaziale Italiana, Rome, Italy\\
$^{21}$Instituto de Estudios Astrof\'{i}sicos, Facultad de Ingenier\'{i}a y Ciencias, Universidad Diego Portales, Santiago, Chile\\
$^{22}$Kavli Institute for Astronomy and Astrophysics, Peking University, Beijing, China\\
$^{23}$Physics Department, Tor Vergata University of Rome, Rome, Italy\\
$^{24}$INAF -- Astronomical Observatory of Rome, Monte Porzio Catone, Italy\\
$^{25}$INFN -- Rome Tor Vergata, Rome, Italy\\
$^{26}$Department of Astronomy, University of Maryland, College Park, MD, USA\\
$^{27}$Department of Physics, SUNY Brockport, Brockport, NY, 14420, USA
}
\def\corrAuthor{Joanna M. Piotrowska}

\def\corrEmail{joannapk@caltech.edu}

\begin{document}
\onecolumn
\firstpage{1}

\title {The High Energy X-ray Probe (HEX-P): Constraining Supermassive Black Hole Growth with Population Spin Measurements} 

\author[\firstAuthorLast ]{\Authors} 
\address{} 
\correspondance{} 

\extraAuth{}

\maketitle

\begin{abstract}

\section{}
Constraining the primary growth channel of supermassive black holes (SMBH)
remains one the most actively debated questions in the context of cosmological
structure formation. Owing to the expected connection between SMBH spin parameter evolution 
and the accretion and merger history of individual black holes, population spin measurements 
offer a rare observational window into the SMBH cosmic growth.
As of today, the most common method for estimating SMBH spin relies on modeling 
the relativistically broaden atomic profiles in the reflection spectrum observed in 
X-rays. In this paper, we study the observational requirements needed to confidently
distinguish between the primary SMBH growth channels, based on their distinct 
spin-mass distributions predicted by the Horizon-AGN cosmological simulation. 
In doing so, we characterize outstanding limitations associated with the existing 
measurements and discuss the landscape of future observational campaigns,
which can be planned and executed with future X-ray observatories. We focus
our attention on the High-Energy X-ray Probe (HEX-P), a~concept probe-class mission aimed 
to serve the high-energy community in the 2030s.

\tiny
 \keyFont{ \section{Keywords:} supermassive black holes, AGN, black hole growth, black hole spin, future X-ray observatories} 
\end{abstract}

\section{Introduction}
\label{sec:intro}

As of today, the existence of supermassive black holes (SMBHs)
residing in the nuclei of galaxies is no longer a subject of scientific 
dispute. With SMBH presence revealed in the orbital motion of stars
in our Galactic centre \citep[e.g.][]{Ghez2008, Genzel2010} 
and direct imaging of matter just outside the Event Horizon \citep{EHT2019, EHT2022}, 
the focus has now shifted
towards understanding the origin and growth of these extreme 
astrophysical objects with masses above $\gtrsim 10^6 \rm M_\odot$.
Decades of extragalactic observations have demonstrated a tight
connection between the properties of supermassive black holes and 
their galactic hosts, now interpreted as black hole-galaxy co-evolution
across cosmic time \citep[see][for a review]{KormendyHo2013}. More
specifically, SMBHs have been established as key players in shaping 
galaxy formation, structure and, most of all, in suppressing 
galactic star formation through a range of highly energetic processes
collectively referred to as Active Galactic Nucleus (AGN) feedback
\citep[e.g.][among many others]{McNamara2000, Bower2006, Terrazas2016,
Henriques2019, Piotrowska2022}. 

Regardless of its exact mode of operation, AGN feedback relies 
on extracting power from the immediate surroundings of supermassive
black holes via accretion. The amount of energy released 
in the process greatly exceeds the threshold required to offset cooling around 
galaxies or to unbind baryons within them, rendering SMBH 
a formidable source of power capable of affecting their galactic hosts 
\citep[see][for in-depth reviews]{Fabian2012, Werner2019}.
At high accretion rates ($\dot{M}_{\rm BH} \gtrsim 0.01 \dot{M}_{\rm Edd}$,
the Eddington accretion rate\footnote{Accretion rate associated with Eddington 
luminosity}), matter surrounding
an SMBH is thought to infall through a geometrically thin, 
optically thick accretion disk \citep{Shakura73, NovikovThorne1973}
in which emitted radiation drives high-velocity 
\textit{quasar} outflows \citep[e.g.][]{Murray1995, King2008, Faucher-Giguere2012}
coupling to the interstellar medium (ISM) 
and further accelerating it at galaxy-wide scales 
\citep[e.g.][]{Feruglio2010, Hopkins2010, Villar-Martin2011, Maiolino2012, 
Cicone2014, Fiore2017}. 
At low accretion rates $\lesssim 10^{-6}\ \dot{M}_{\rm Edd}$, 
the inflow forms a~geometrically thick, optically thin accretion disk 
\citep[e.g.][]{Yuan2014, Giustini2019} and launches \textit{relativistic jets} which
deposit energy within the circumgalactic medium (CGM) 
at large distances away from the galactic host
\citep[e.g.][]{McNamara2000, Birzan2004, Hlavacek-Larrondo2012,  
Hlavacek-Larrondo2015, Werner2019}. 
Across all flavours of AGN feedback processes, the efficiency
with which power is extracted from the accretion flow depends
on the angular momentum of the SMBH and spans over an order 
of magnitude in range \citep[e.g.][]{Thorne74, Penna2010, Avara2016, Liska2019}, 
further broadening the range of impact AGN can have on their host galaxies.

Although abundant observational evidence exists for the 
impact of supermassive black holes on their surrounding galaxies, 
relatively little is known about SMBH origin and growth across cosmic time. 
Understanding how these black holes form and reach their 
impressive masses is particularly important both
in the context of galaxy evolution and recent James Webb Space Telescope 
\citep[JWST,][]{JWST, JWST2} observations, 
which report black holes with masses $M_{\rm BH} > 10^6 \rm M_\odot$
as early as $z=10.6$ \citep{Harikane2023, Maiolino2023b, Maiolino2023a}. Since
cosmic growth via accretion of gas and SMBH-SMBH mergers 
occurs on timescales beyond direct 
human observation, one would, ideally, like to characterise
supermassive black hole growth histories using alternative
observables. An example of such a proxy is black hole 
angular momentum, $\vec{J}$, which changes its orientation and magnitude 
in response to different physical processes increasing \mbh. Hence, 
by \textit{measuring the magnitude of supermassive black hole angular momentum $J$,} 
usually expressed in terms of the dimensionless 
spin parameter ($a^\ast \equiv J c / GM_{\rm BH}^2$, where $c$ is the speed
of light and $G$ is the gravitational constant), 
\textit{one can hope to extract information about the past record of its growth.}
In the AGN, spin parameter can be estimated by modelling coronal X-ray radiation 
reprocessed (or `reflected') by the accretion disk \citep[e.g.][]{Fabian89, kdblur}. 
Because the reflected signal depends on the position of the innermost 
edge of the accretion disk, this relationship can be directly applied
to determine \spin\ in nearby SMBH (see \ref{sec:spin} for an overview
of this approach). 

As the black hole increases its mass via accretion of surrounding 
material, its spin changes owing to angular momentum transfer 
from the accretion flow.
If the accreting material settles into a~prograde\footnote{In a prograde 
disk, the accreting material and the black hole are both rotating in the 
same direction} disk, inflow of angular momentum aligned with that of the 
black hole leads to its efficient spin-up \citep{Bardeen1970, Moderski1996, Moderski1998}.
In contrast, chaotic accretion of matter with randomly oriented angular momentum 
decreases the spin magnitude, ultimately driving it towards $a^\ast \sim 0$ 
over sufficiently long times \citep[e.g.][]{Berti08, King2008, Dotti2013}. In the
presence of magnetic fields, even in the case of ordered inflow through 
aligned accretion disks, spin can also be reduced by energy extraction 
via the Blandford-Znajek process \citep{Blandford1977}. 
This jet launching mechanism has been shown to drain black hole angular 
momentum in magnetically arrested disks (MAD) 
in general-relativistic magneto-hydrodynamical (GRMHD) simulations 
of SMBH accretion \citep[e.g.][]{Tchekhovskoy2011, Tchekhovskoy2012, 
McKinney2012, Narayan2022, Curd2023, Lowell2023}. Finally, mergers
of supermassive black hole binaries leave behind remnants which can be 
either spun up or down with respect to their progenitors, contingent on individual 
spin parameters upon coalescence \citep[e.g.][]{Kesden2008, Rezzolla2008,
Barausse2009, Tichy2008, Healy2014, Hofmann2016}.

Depending on the relative contribution of these processes across cosmic
time, one would expect different growth histories of SMBH to leave an 
imprint on the measured spin parameter. In the cosmological context, 
this expectation was first explored in semi-analytic models (SAMs) 
- taking advantage of their low computational cost, several studies
have now used SAMs to make spin population predictions for
different SMBH accretion and coalescence scenarios. 
\cite{Berti08} demonstrated that prolonged episodes of coherent 
accretion produce SMBH populations spinning at near-maximal rates, 
while chaotic infall of matter and mergers
force the dimensionless spin parameter towards $a^\ast \sim 0$. \cite{Dotti2013}
generalised the chaotic accretion paradigm and found that SMBHs are not 
spun down efficiently when the distribution of accreted angular momenta 
is not isotropic, allowing black holes to maintain stable high spin values
for even modest degrees of anisotropy. By linking the orientation of accreted
angular momentum with that of the galactic host, \cite{Sesana2014} further 
showed that spin parameter critically depends on the dynamics of the host and 
that SMBH residing in spiral galaxies tend to spin fast, biasing the observable
samples towards high $a^\ast$ values. 

Moving beyond the idealised semi-analytic approach, spin parameter modelling
in hydrodynamical cosmological simulations only recently became
an area of active development \citep{Dubois2014a, Fiacconi18, 
Bustamante19, Talbot2021, Talbot2022}. As of today, 
there exist three major simulation suites which trace spin parameter
in statistical samples of SMBHs. The \textit{NewHorizon cosmological 
simulation} \citep{Dubois2021}, on-the-fly spin evolution coupled 
to AGN feedback prescription in a $\rm (16\ Mpc)^3$ volume;
simulated down to $z=0.25$ at a maximum spatial resolution of 34~pc.
The \textit{\cite{Bustamante19} simulation suite}, on-the-fly 
spin evolution implemented in the moving-mesh code \textsc{AREPO} \citep{Springel2010}; 
a cosmological volume of $(\sim \rm 37 Mpc)^3$ with a spatial resolution 
of $\sim 1\ \rm kpc$; evolved to $z=0$ with the IllustrisTNG AGN feedback 
model \citealt{Weinberger2017}. Finally, the \textit{Horizon-AGN cosmological
simulation} \citep{Dubois2014b}, in which spin evolution is computed via post-processing 
of the completed Horizon-AGN run \citealt{Dubois2014c}, 
followed
down to $z=0$ in a $(\rm \sim 147\ Mpc)^3$ volume at maximum
spatial resolution of $\sim \rm 1\ kpc$. All three suites produce 
realistic populations of SMBH and yield \spin\ population statistics compatible 
with those observed in the local Universe. With limited sample sizes and 
generous uncertainty on individual measurements, currently available 
observations are not constraining enough to indicate preference 
for any particular model implementation. In the future, increased
measurement precision and improved statistics in the \spinmass\ plane  
will be of critical importance for new generations of on-the-fly spin
evolution models: with improved calibration targets for subgrid prescriptions,
different hydrodynamical cosmological models are likely to further converge 
over time, allowing us to use \spin\ constraints as an interpretative tool, 
as opposed to means of discriminating among modelling approaches.

In this study, we show how future observing programs targeting 
statistical samples of SMBH spins and masses can be designed to 
best discriminate between different growth histories of 
supermassive black holes. We choose to focus on the Horizon-AGN
cosmological simulation to take advantage of its excellent SMBH
statistics delivered by the substantial simulation volume.
Treating the simulation suite as a case study, we estimate sample sizes and measurement 
precision required to differentiate between accretion- and 
merger-dominated SMBH growth. We then discuss these requirements
in the context of the High-Energy X-ray Probe (\hexp; \textbf{[Madsen et al.\ 2023]}) 
-- a~probe-class mission concept which offers sensitive broad-band 
coverage (0.2--80\,keV) of the X-ray spectrum with exceptional spectral, 
timing and angular capabilities, featuring two high-energy telescopes 
(HET) that focus hard X-rays, and soft X-ray coverage with a~single 
low-energy telescope (LET). Taking into account the \hexp\ instrument design,
we explicitly demonstrate the potential for this probe class mission
to deliver SMBH spin parameter measurements necessary for constraining 
supermassive black hole growth in the observable Universe.

In Section~\ref{sec:cosmo-sims} we discuss the challenges 
associated with spin modelling in cosmological hydrodynamical simulations 
and briefly describe the Horizon-AGN suite. 
In Section~\ref{sec:smbh-observations} we provide an overview 
of reflection spectroscopy as a tool for measuring \spin, followed
by a discussion on current constraints and their comparison
with the Horizon-AGN cosmological model in Section~\ref{sec:current-constraints}.
In Section~\ref{sec:need-for-study} we advocate for a systematic
study of the \spinmass\ plane with future observing programs 
and in Section~\ref{sec:bass} we describe a~sample of AGN selected
from the BAT AGN Spectroscopic Survey appropriate for such study. In Section~\ref{sec:sample-reqs}
we determine minimum sample requirements for differentiating 
between accretion- and merger-dominated SMBH growth, followed by 
\hexp\ spin parameter recovery simulation in Section~\ref{sec:spectral-sims}.
In Section~\ref{sec:job-done} we demonstrate the potential for 
\hexp\ to characterise SMBH growth histories through population
spin measurements and present our final remarks in Section~\ref{sec:conclusions}.

\section{SMBH spin in cosmological hydrodynamical simulations}
\label{sec:cosmo-sims}

Tracing the change in supermassive black hole (SMBH) spin in the context of
large-scale evolution of the Universe is an extraordinarily complex problem
spanning a broad range of spatial and temporal scales. Transfer of angular momentum
via accretion in disks occurs on size scales between $\sim 1$ and a~few $10^2$
gravitational radii $R_G=GM_{\rm BH}/c^2$ \citep[e.g.][]{Jiang2017,
Cackett2018, Fausnaugh2016, Edelson2019, Guo2022, Homayouni2022}, 
which then couples to gas inflows from host galaxy scales of several hundred 
pc \citep[e.g.][]{Garcia-Burillo2005, Hopkins2010, Alexander2012, Wong2011, 
Wong2014, Russell2015, Russell2018}, 
ultimately fuelled by gas accretion from within the cosmic web at the 
intergalactic scales of hundreds of kpc 
\citep[e.g.][]{Sancisi2008, Putman2012, Tumlison2017}. 
At the same time, gas flows at all scales are affected
by the non-trivial interplay of feedback processes from both 
stars \citep[e.g.][]{Katz1996, Cole2000, Keres2005, Hopkins2014} 
and active galactic nuclei, AGN \citep[e.g.][]{Bower2006, Croton2006, Fabian2012, KormendyHo2013}, 
disk instabilities (e.g.~\citealt{Schwarz1981}; \citealt{Athanassoula1983}; 
\citealt{Kuijken1995}; \citealt{Debattista2006})
and galaxy-galaxy interactions (e.g.~\citealt{Toomre1972}; \citealt{Sanders1988},
\citealt{Barnes1992}; \citealt{DiMatteo2005}; \citealt{Springel2005};
\citealt{Hopkins2006}) 
all of which, ideally, need to be taken into account in a comprehensive modelling of 
SMBH spin evolution. The dynamic range in size scales alone renders it computationally 
unfeasible to directly follow SMBH evolution
even with current state-of-the-art hardware and numerical methods 
\citep[see][for reviews of the current state of the field]{Vogelsberger2020, Crain2023}. 
Hence, to overcome these limitations, cosmological simulations 
replace missing baryonic physics with 
\textit{subgrid}\footnote{Capturing physics at scales 
smaller than those allowed by numerical resolution via semi-analytic prescriptions
coupled to quantities directly resolved in the simulation. See \cite{Vogelsberger2020}
and \cite{Crain2023} for comprehensive reviews of the methodology.} 
prescriptions - semi-analytic models relying
on astrophysical scaling relations to predict large scale hydrodynamic effects 
of unresolved AGN accretion, spin evolution and feedback. 

There currently exist three cosmological 
hydrodynamical simulations which either model spin on-the-fly 
\citep{Bustamante19, Dubois2021}
or calculate its evolution in post-processing \citep{Dubois2014b}
(see Section~\ref{sec:intro}). 
Although these three studies differ significantly in their subgrid prescriptions,
in addition to spanning different cosmological volumes and final simulation redshift, 
they all deliver SMBH spin populations broadly consistent with the
currently available observational constraints. As spin modelling 
in full cosmological context is still in its infancy, the agreement between 
both models themselves and with the observable Universe is likely to improve,
once better constraints are available for subgrid parameter tuning.
Thus, the study presented in this article does not focus on differentiating
among currently available models, but rather on demonstrating the capability 
of \hexp\ to identify different SMBH growth channels within a~single 
realisation of a~sophisticated full-physics model of the Universe.

\subsection{Horizon-AGN cosmological model}
\label{sec:horizonAGN}

In our study we make use of the Horizon-AGN cosmological simulation \citep{Dubois2014c}, 
post-processed to trace SMBH spin evolution in response to local hydrodynamics of gas, 
accretion and SMBH mergers \citep{Dubois2014b}. Our choice of the simulation suite is 
motivated by its substantial volume of $(\sim 100 \rm cMpc)^3$, generous 
statistics and great success in reproducing realistic SMBH and galaxy samples across 
cosmic time \citep{Dubois2016, Volonteri2016, Beckmann2017, Kaviraj2017}. 
Most importantly, 
however, accretion-dominated and  accretion \& merger growth histories of supermassive 
black holes in Horizon-AGN have been shown to result in distinct spin parameter 
distributions \citep{Beckmann2023}, hence the suite offers a promising opportunity 
for studying the SMBH growth signal with \hexp\ reflection spectroscopy. 
The full Horizon-AGN suite, together with its spin parameter evolution model
are described in detail in \cite{Dubois2014c} and \cite{Dubois2014b} and here we 
only provide a~brief overview of SMBH treatment in the simulation.

Horizon-AGN is a suite of cosmological, hydrodynamical simulations performed
with the \textsc{RAMSES} code \citep{Teyssier2002} for $\rm \Lambda CDM$ cosmology
with cosmological parameter values estimated in WMAP-7 observations 
($H_0=70.4 \rm km   s^{-1} Mpc^{-1}$, $\Omega_m = 0.272$, $\Omega_b=0.045$,
$\Omega_\Lambda = 0.728$ and $\sigma_8=0.81$; \citealt{Komatsu2011}). 
The hydrodynamics computed on an adaptively refined Cartesian grid is coupled
to subgrid prescriptions for baryonic interactions, 
including gas cooling, star formation, 
black hole accretion and feedback processes described in detail in \cite{Dubois2014c}. 
The simulation only includes black holes
of the supermassive kind, seeding them in $z>1.5$  gas cells once these exceed
a star formation threshold of $n_0=0.1 \rm H\, cm^{-1}$ in density. Once seeded, 
the SMBH then accrete via a boosted Bondi-Hoyle-Lyttleton prescription 
($\dot{M}_{\rm BH} = 4 \pi \alpha G^2 M_{\rm BH}^2 \bar{\rho}/(\bar{c}_s^2 + \bar{u}^2)^{3/2}$,
where $\bar{u}$, $\bar{\rho}$ and $\bar{c}_s$ are gas velocity, density and speed
of sound averaged in the SMBH vicinity, and $G$ is the gravitational constant; 
\citealt{Hoyle1939}; \citealt{Bondi1944}; \citealt{Bondi1952}) 
capped at the Eddington rate. A fraction $\epsilon_r=0.1$ of the accreted 
rest-mass energy is then released as AGN feedback with 
$\dot{E}_{\rm AGN} = \epsilon_r \dot{M}_{\rm BH} c^2$, 
where $c$ is the speed of light, and the 
fraction of power coupled to gas is controlled by the accretion rate-dependent
feedback mode. 
For Eddington ratios $f_{\rm Edd} \equiv \dot{M}_{\rm BH}/\dot{M}_{\rm Edd} > 0.01$, 
the AGN is in the `quasar' mode and isotropically injects $0.15\ \dot{E}_{\rm BH}$ 
as thermal energy around the SMBH particle. For $f_{\rm Edd} < 0.01$
the AGN switches to `radio' mode feedback, releasing all $\dot{E}_{\rm BH}$
in biconical outflows. In a post-processing procedure introduced in \cite{Dubois2014b},
SMBH are assumed to begin with a spin of $a^\ast=0$, which subsequently 
evolves in response to accretion of gas from the hydrodynamical grid and 
SMBH-SMBH mergers across cosmic time. The spin evolution model 
combines semi-analytic considerations and calibrations extracted from GRMHD simulations
to model the two phenomena, including spin evolution in misaligned disks
via the Bardeen-Peterson effect. The prescription, however, does not model 
SMBH spin-down due to jets via the Blandford-Znajek mechanism.

\section{Observations of black holes}
\label{sec:smbh-observations}

Astrophysical black holes are unique objects of great interest as they represent the ultimate test of general
relativity as a~theory for gravity. Despite their exotic phenomenology, black holes are also objects of 
outstanding simplicity; owing to the no-hair theorem \citep{Israel1967, Israel1968, Carter1971}, they are fully
described by only three physical quantities: mass, angular momentum (or spin), and electric charge. Furthermore, charge
is quickly neutralized in any realistic astrophysical environment \citep{Michel1972, Wald1984, Novikov1989, Treves1999, bambi2017}, 
leaving just mass and spin as the fundamental parameters to be constrained by observations.

\subsection{Mass}

Of the two properties describing a~black hole, mass is likely the most straightforward to 
measure (both for stellar and supermassive black holes; SMBHs). Indeed, a range of
techniques has now been developed for SMBH mass estimation, appropriate for both local and distant sources.
For a~relatively nearby system (such as the SMBH in the center of our Galaxy), 
it is possible to resolve the motion of individual stars and model the underlying 
gravitational potential to estimate its associated black hole mass (e.g.~\citealt{Ghez1998}; 
 \citealt{Ghez2008}; \citealt{Genzel2010}; \citealt{GRAVITY2018}). Owing to extremely 
high resolution requirements, such studies can only be undertaken within the Milky Way, 
while extragalactic \mbh\ measurements are instead forced to rely on stellar, gas and 
maser kinematics (e.g.~\citealt{Kormendy1998}; \citealt{Magorrian1998}; 
 \citealt{Gebhardt2000}). 
 
One of the most robust methods for 
measuring \mbh\ for active SMBHs in other galaxies involves reverberation mapping of 
broad emission lines \citep{Blandford1982, Peterson1993}. In this approach, velocity 
dispersion of gas in the broad line region (BLR) is combined with a~measurement 
of the time delay between variations in the continuum and the line emission to 
infer the dynamics of the accretion disks 
\citep[see e.g.][for a~description of the methodology]{Peterson2004}. 
The mass estimate relies on the proportionality\footnote{The proportionality 
constant, known as the virial factor, is challenging to determine for individual 
objects owing to the unknown geometry of the BLR gas \citep[e.g.][]{Brewer2011, Pancoast2014}} $M_{\rm BH} \propto R_{\rm BLR} ({\Delta}V)^2$ 
between SMBH mass, the line-of-sight velocity of BLR gas (${\Delta}V$, encoded in 
emission line profiles) and the distance between the black hole and the BLR ($R_{\rm BLR}$,
inferred from the time lag measurement). Since observations of this kind require long
exposure times at prime facilities, the current sample of \mbh\ measured via 
reverberation contains only $\sim 100$ objects in total (e.g.~\citealt{Woo2015}). 

Because of the limited number of direct SMBH mass measurements, large statistical 
studies commonly rely on empirical calibrations to estimate \mbh\ 
from the measured properties of their galactic hosts. Among these, the relationship
between stellar velocity dispersion in the galactic bulge and SMBH mass measured 
via reverberation mapping, the $M_{\rm BH} - \sigma_\ast$ relation, boasts 
the least scatter, yielding a~systematic uncertainty of $\sim 0.3-0.5$~dex 
in $\log(M_{\rm BH})$ \citep{Ferrarese2000, Hopkins2011, MM13, Saglia2016}. 
Although subject to significant scatter at present, these calibrations will
improve with the arrival of future large multi-epoch observation programs
like the Black Hole Mapper \citep{BHmapper2019}. The planned $\sim 1000$ 
optical reverberation mapping \mbh\ estimates will significantly
increase the robustness and precision of the $M_{\rm BH} - \sigma_\ast$ estimation.
Other techniques have also been recently proposed to measure \mbh\ using X-ray variability, 
such as X-ray reverberation \citep{Alston2020} and X-ray spectral-timing \citep{Ponti2012, Ingram2022}, 
both of which can be model dependent and carry systematic uncertainties that still need 
to be understood.

\subsection{Spin} 
\label{sec:spin}

The estimation of black hole spin, on the other hand, is a somewhat more difficult task, as it mainly
impacts the environment very close to the black hole. For example, the black hole's angular momentum determines 
the radius of the innermost stable circular orbit, ISCO ($R_{\rm{ISCO}}$, which varies from 9\,$R_{\rm{G}}$
for a~maximal retrograde spin to  $\sim$1.2\,$R_{\rm{G}}$ for a~maximal prograde
spin of $a^\ast = 0.998$, where $R_{\rm{G}} = GM_{\rm{BH}}/c^{2}$ is the gravitational radius;
\citealt{Bardeen72, Thorne74}). In principle, the spin of a~black hole can be
constrained through a~variety of methods, most of which involve determining
$R_{\rm{ISCO}}$, but for AGN the most reliable method currently available comes via
the characterization of the relativistic reflection from the innermost accretion
disk (sometimes referred to as the iron-line method; see \citealt{Reynolds21rev}for a recent review).

\begin{figure*}
\begin{center}
\hspace*{-0.4cm}
\rotatebox{0}{
{\includegraphics[width=155pt,trim={0cm -0.5cm 0cm 0.5cm},clip]{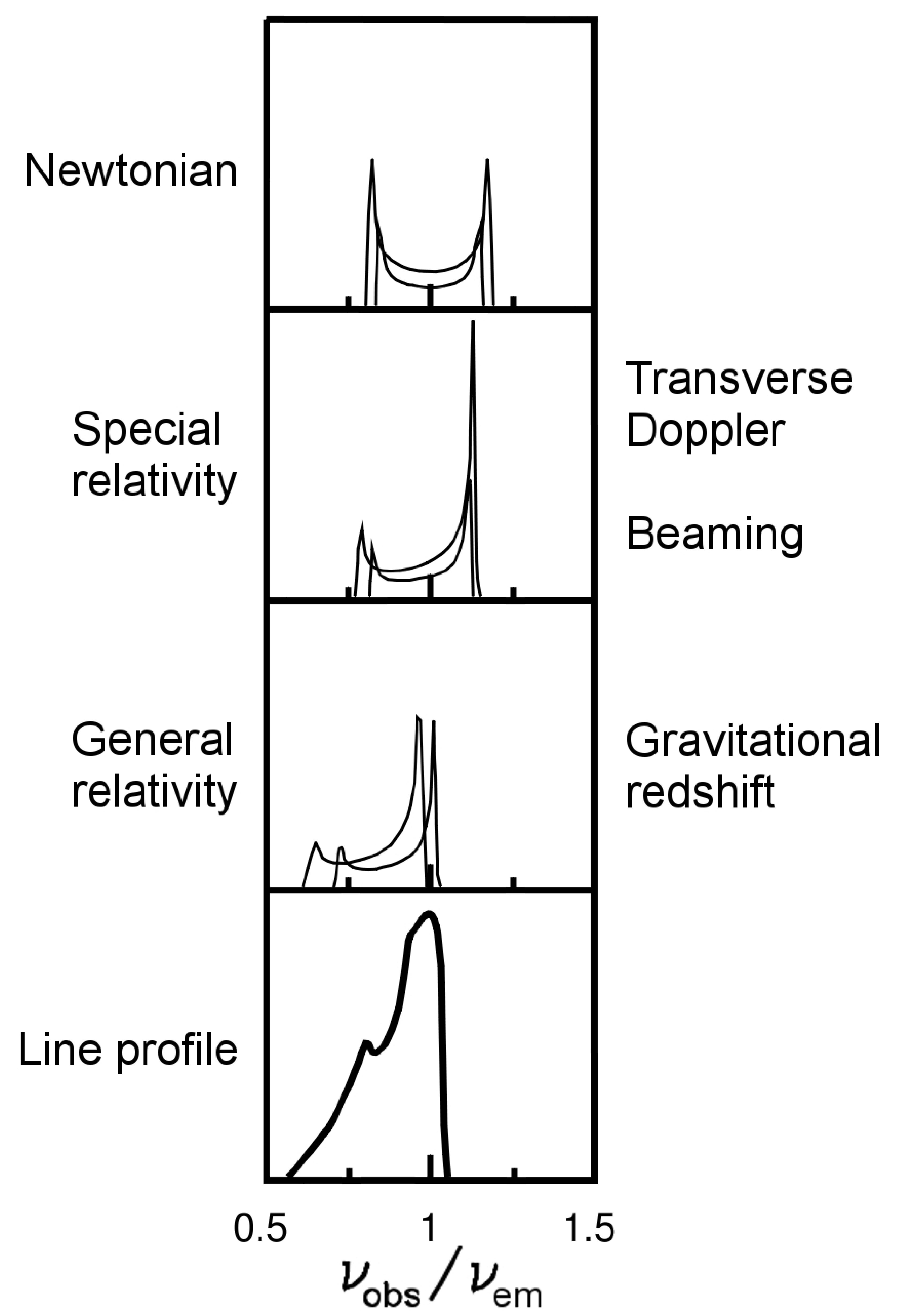}}
}
\hspace*{0.6cm}
\rotatebox{0}{
{\includegraphics[width=300pt]{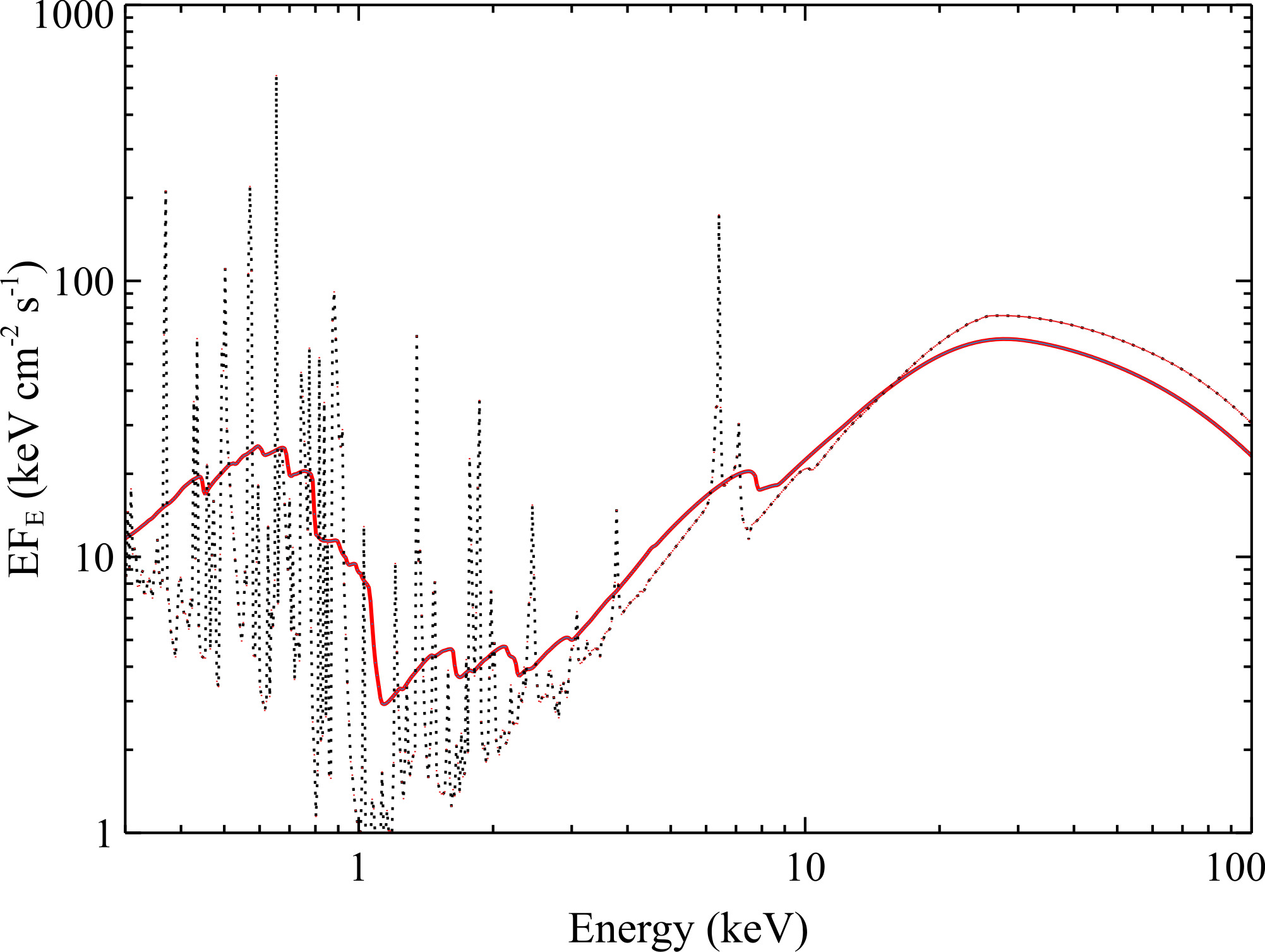}}
}
\end{center}
\vspace*{-0.3cm}
\caption{
\textit{Left panel:} The broadening experienced by a~narrow emission line from
an accretion disk around a~black hole (from \citealt{Fabian00}). The top three
panels show the effects of Newtonian gravity, special relativity and general
relativity, respectively, on the line profiles for two example radii
within the disk, while the bottom panel shows the broadened and skewed
`diskline' profile expected when these combined effects are integrated over the
full radial profile of the disk.
\textit{Right panel:} An example of these relativistic effects applied to a
typical reflection spectrum from moderately ionised material (as may be expected
for the accretion disk in an AGN). Here we use the \textsc{xillver} reflection
model for the rest-frame reflection spectrum (dotted line; \citealt{xillver})
and apply the relativistic blurring with the \textsc{relconv} model (solid line;
\citealt{relconv}).}
\label{fig:blur}
\end{figure*}

For moderately high accretion rates most of the infalling material is expected to
flow through a~thin, optically-thick accretion disk (\citealt{Shakura73}). Some of
the thermal emission from the disk, which peaks in the UV for most AGN, is Compton
up-scattered into a~much higher energy continuum by a~`corona' of hot electrons
($kT_{\rm{e}} \sim 100$\,keV; \citealt{Fabian15, Balokovic20}), which is the primary
source of X-ray emission in AGN\footnote{The precise nature of the corona~is still
poorly understood, but is also an area~of AGN physics that \textit{HEX-P} will make
significant contributions to (\textbf{[Kammoun et al. 2023]})}. This X-ray emission
re-irradiates the surface of the disk, producing a~further `reflected' emission
component that contains a~diverse range of atomic spectral features both in emission 
and absorption. Among these, the inner-shell transitions from iron ions - predominantly the
K$\alpha$ emission line at 6.4--6.97\,keV (depending on the ionisation state) -
together with a strong absorption K-edge around $7-8$\,keV, are typically the most prominent. 
Additionally, reflection spectra exhibit a~characteristic high-energy continuum, peaking at
$\sim$30\,keV, often referred to as the `Compton hump' (\citealt{George91, reflion,
xillver}). Although the emission lines are narrow in the frame of the disk, by the
time the material in the disk approaches the ISCO it is orbiting at relativistic
speeds, and the gravitational redshift is also very strong. The combination of
these effects serves to broaden and skew the emission lines from our perspective as
a~distant, external observer, resulting in a~characteristic `diskline' emission
profile (\citealt{Fabian89, kdblur, kerrconv, relconv}; see Figure \ref{fig:blur}).
Provided that the disk extends all the way into the innermost stable circular
orbit, also expected at moderately high accretion rates, then characterising the
relativistic blurring gives a~measurement of $R_{\rm{ISCO}}$, and thus of $a^*$.
These distortions are most often discussed in the context of the iron emission
line specifically, as it is typically the strongest spectroscopically isolated emission 
line in the reflection spectrum. While the iron emission provides the cleanest
view of the broadened line profile, in reality, these relativistic effects
impact the whole reflection spectrum (see also Figure \ref{fig:blur}).

This approach to measuring spin is particularly powerful as it can be applied to
black holes of all masses, not just the SMBHs powering AGN (e.g.
\citealt{Walton12xrbAGN}). Indeed, relativistically broadened Fe K lines have been
observed in the spectra~of a~large fraction of AGN with high signal-to-noise (S/N)
X-ray spectra~(e.g. \citealt{Tanaka95, Miniutti07iras, FabZog09, dLCPerez10,
Brenneman11, Nardini11, Gallo11mrk79, Parker14mrk, Ricci14, Wilkins15, Jiang18iras,
Walton19ufo}) as well as most well-studied BH X-ray binaries (e.g.
\citealt{Miniutti04, Miller09, Reis09spin, Duro11, King14, Garcia15gx, Walton17v404,
Xu18, Kara19, Tao19grs, Jiang22, Draghis23}; see also \textbf{[Connors et al. 2023]}
for more discussion on BH X-ray binaries with \textit{HEX-P}). Broadband X-ray
spectroscopy is particularly critical for properly characterising the reflected
emission: its key features span the entire observable X-ray band (a
diverse set of emission lines from lighter elements at energies $\lesssim 2 \rm \, keV$,
the iron emission line at $\sim$6--7\,keV and the Compton hump at $\sim$30\,keV)
and their proper modelling requires an accurate characterisation of the continuum
across the entire broadband range.
The \nustar\ era~has therefore marked a~period of excellent recent progress in this
field, finally providing high S/N spectroscopy up to $\sim$80\,keV for AGN.
\nustar\ observations have unambiguously revealed the high-energy reflected continuum
associated with the broad iron emission, and have been particularly powerful when
paired with simultaneous lower-energy coverage from e.g. \textit{XMM-Newton},
confirming our ability to measure spin via~reflection spectroscopy (e.g.
\citealt{Risaliti13nat, Walton14, Marinucci14, Buisson18, Garcia19, Chamani20,
Wilkins22}). As we further show in Section~\ref{sec:current-constraints}, however,
these currently available samples of measurements with
X-ray reflection spectroscopy are still not sufficient for characterising 
SMBH growth histories in the observable Universe.

\section{Current Constraints: the Spin--Mass Plane}
\label{sec:current-constraints}

The most recent systematic compilations of SMBH spins and masses from the literature are
presented in \cite{Reynolds21rev} and \cite{Bambi21rev}, although there have been
a~few more measurements made since the publication of these reviews
(\citealt{Walton21spin, Mallick22, SiskReynes22}), resulting in a~current sample
size of 46 SMBHs with at least initial spin constraints. All but three of
these sources are from the local Universe ($z \leq 0.3$, and even then most have
$z \leq 0.1$); the two highest redshift constraints to date come from
strongly-lensed quasars ($z = 0.66$ and $z = 1.7$; \citealt{Reis14nat,
Reynolds14}).

\begin{figure}
    \centering
    \includegraphics[width=0.8\textwidth]{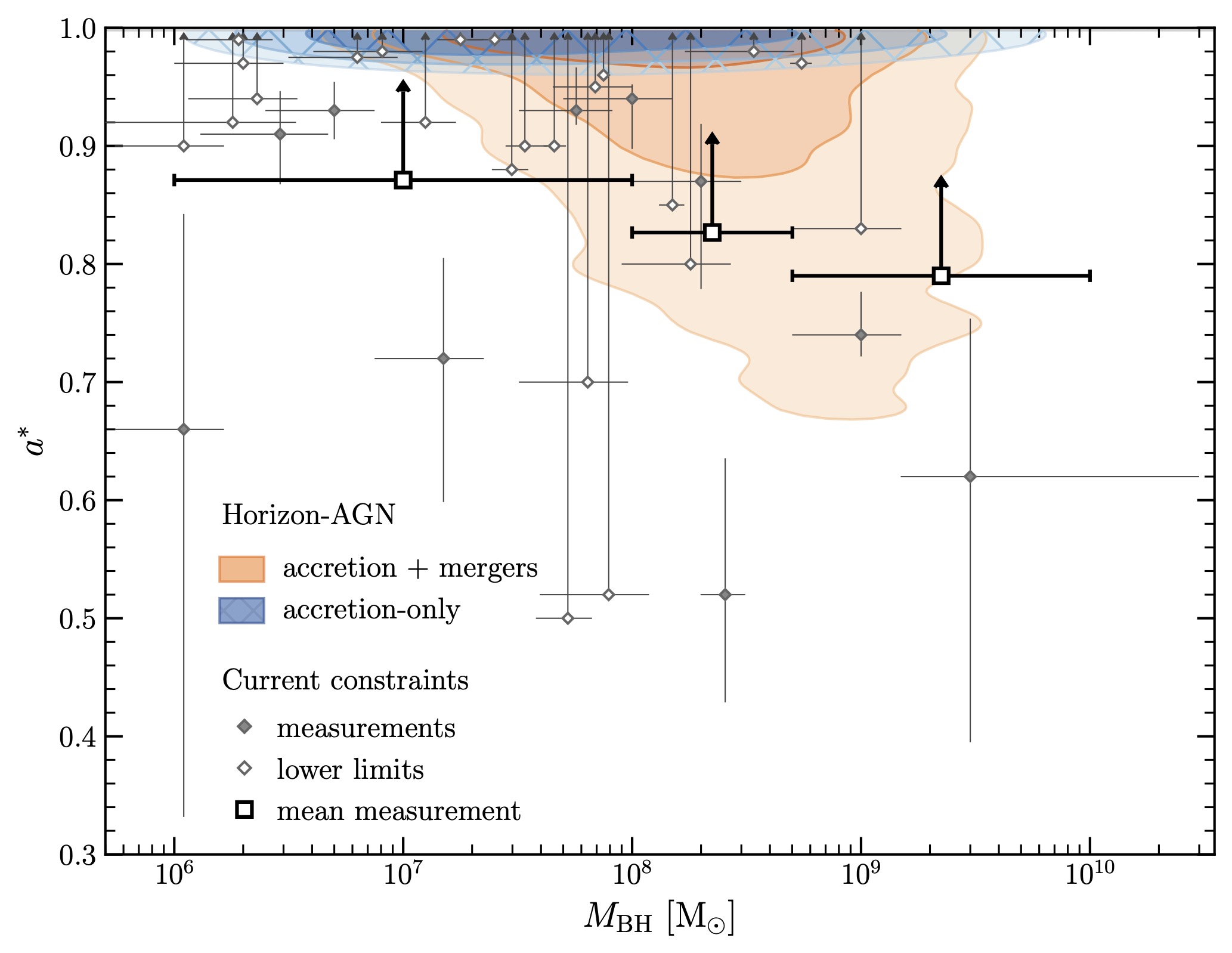}
    \caption{Current observational constraints in the \spinmass\ plane 
    (open and filled grey diamonds), compared against the locus occupied by 
    the HorizonAGN cosmological simulation (logarithmically spaced contours) 
    for black holes with $M_{\rm{BH}} \geq 10^{6}$\,\msun\ (the range
    covered by Horizon-AGN). Error bars associated with filled diamonds
    indicate $1\sigma$ uncertainty on \spin\ measurement. 
    Blue, hatched: SMBHs which acquired less than 10\% of their 
    mass in mergers (i.e. accretion-only growth),
    orange: SMBH with $>10\%$ mass grown in mergers (i.e. accretion + mergers growth). 
    Both contours account for the
    radiative efficiency - spin bias in flux-limited AGN samples. Existing
    constraints indicate a~decreasing trend in \spin\ with \mbh, as shown by lower limits
    on mean \spin\ values in \mbh\ bins (black open squares), however are not
    able to differentiate between SMBH growth scenarios (orange vs. blue contours).}
    \label{fig:tongues}
\end{figure}

Fig.~\ref{fig:tongues} shows the current constraints in the spin-mass plane 
for SMBH with masses above $10^6 \rm M_\odot$. Grey filled diamonds indicate
measurements with $1\sigma$ error bars\footnote{Measurements reported 
in the literature commonly
quoute 90\% confidence intervals for spin parameter measurements. In order
to translate these values to $1\sigma$ estimates, we divide the confidence
interval by a factor of 1.64, necessarily assuming that the current constrains 
follow split normal distributions}, 
while open diamonds mark lower limits on~\spin.
Many of the measurements show evidence for rapidly rotating black holes,
particularly for masses in the range $M_{\rm{BH}} \sim 10^{6} - 10^{7}$\,\msun.
There are hints in the current data~of trends towards lower spins, or at least an
increase in the spread of spin measurements, at higher masses particularly 
for $M_{\rm{BH}} > 10^{8}$\,\msun. In order to better capture these trends 
we further calculate mean spin parameter values in three bins of 
SMBH mass between $10^6$ and 
$10^{10}\ \rm M_\odot$\footnote{Our choice of binning scheme is discussed 
in Section~\ref{sec:sim-setup}}, marked with open black squares. 
In the mean calculation we treat all available constraints as uncensored measurements
and hence arrive at lower limit estimates in each bin, which indicate 
a~tentative decrease in \spin\ with increasing \mbh. We note,
however, that there are still only a few measurements in the high-mass regime, 
and the uncertainties on those are also still quite poor. Furthermore,
although a few sample-based efforts to constrain SMBH spin 
via~systematic relativistic reflection analyses exist
\citep[e.g.][]{Walton13spin, Mallick22}, the majority of these measurements come
from independent analyses of individual sources. As such, they are quite
heterogeneous with regards to the exact reflection and relativistic blurring
models used, the precise assumptions adopted in the use of these models, and
the energy range over which the analysis has been performed; moreover, \nustar\ has
only contributed to $\sim$half of these measurements, so not all of them are
based on high S/N broadband X-ray spectroscopy.

Perhaps one of the most important caveats associated with current spin 
measurement constraints is the target selection. In a~collection of 
independent observations, as opposed to a~survey campaign, it is challenging
to properly control for the sample selection function and, by extension, 
observational biases present in the combined dataset. Among these, a~potential 
bias towards high \spin\ values resulting from the expected spin-dependence of
the radiative efficiency (\radeff) in a~relativistic thin-disk solution 
\citep{NovikovThorne1973} is frequently discussed in the literature as at
least part of the reason for the observed prevalence of near-maximal spin
values \citep[e.g.][]{Brenneman11, Reynolds2012}. Under the simplifying
assumption of a~uniform distribution of AGN in the local Universe, the
probability of observing a~given spin value \spin\ scales with \radeff\ to
power $\flatfrac{3}{2}$ for sources with equal accretion rates in flux-limited 
surveys (see Appendix for further details). The superlinear scaling combined 
with steep increase of \radeff\ at $a^\ast > 0.8$ can have profound 
implications for inferred spin population statistics in observations.
 
\begin{figure}[t]
\centering
        \includegraphics[width=\textwidth]{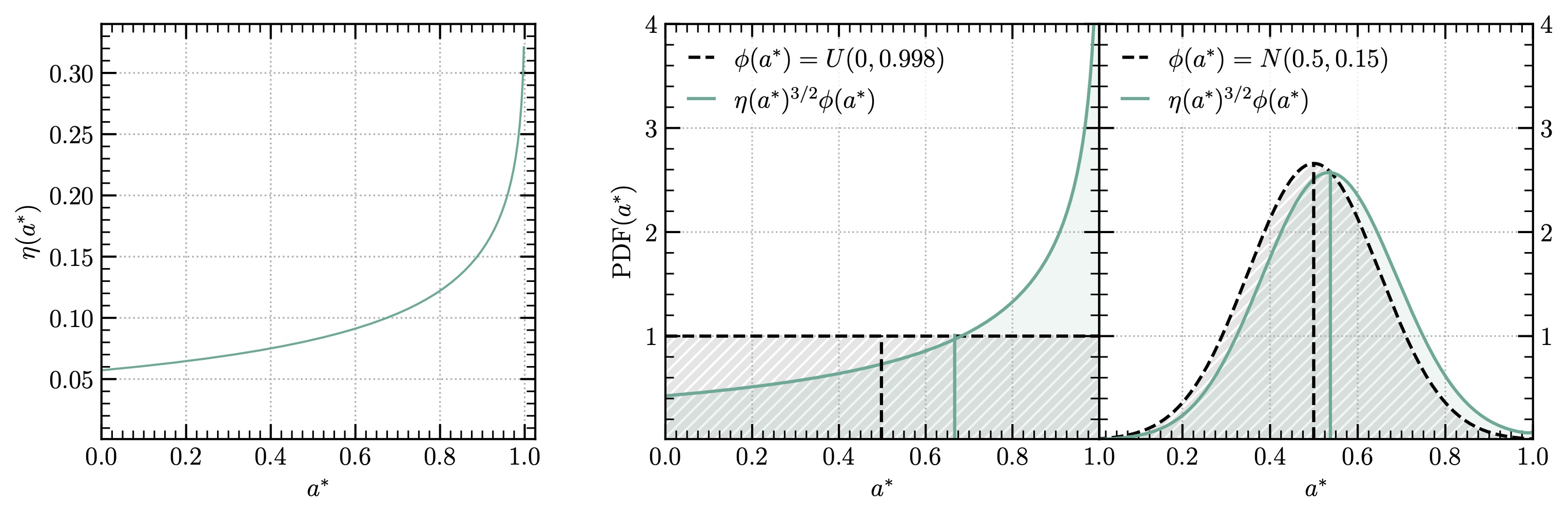}
        \caption{\textit{Left panel}: efficiency $\eta(a)$ as a~function of spin
        parameter $a$ for a~thin disk solution \citep{NovikovThorne1973}.
        \textit{Right panels}: the effect of spin-dependence of $\eta$ on the measurement
        spin population statistics, illustrated with changes to the probability
        distribution function (PDF, shaded regions) and its associated mean spin 
        parameter value (vertical lines). Grey hatched regions in the middle and rightmost panels
        correspond to `raw' PDFs for a~uniform distribution $a^\ast \sim U(0, 0.998)$ and 
        a normal distribution centered on $a^\ast=0.5$ with $\sigma=0.15$, respectively,
        while their coloured counterparts to PDFs corrected for the $\eta-a^\ast$ bias.
        The shift between dashed and solid vertical lines indicates the change in associated
        mean \spin\ value between raw (dashed) and $\eta$-bias corrected distributions (solid). 
        The magnitude of bias critically depends on the true underlying spin parameter
        PDF.}
        \label{fig:eta-bias}
\end{figure}

Fig.~\ref{fig:eta-bias} shows the change in radiative efficiency with spin in 
geometrically thin relativistic disks (left panel) together with the potential
impact the $\eta-a^\ast$ bias can have on the observed spin parameter distributions
(two panels on the right). Using a~uniform distribution of $a^\ast \sim U(0,0.998)$ 
and a normal distribution $a^\ast \sim N(0.5, 0.15)$ as examples, the two panels 
on the right demonstrate how a~probability distribution function (PDF) of the 
expected measurements is distorted towards high spin values, shifting the 
expected mean observed spin values towards $a^\ast \approx 0.67$ and 
$a^\ast \approx 0.54$ for the two respective input distributions. 
As demonstrated in the figure, the impact of the $\eta-a^\ast$ bias is critically dependent on the
underlying true spin parameter distribution. Therefore, the strength of this effect will
differ from our test cases for realistic samples of AGN observations.

\subsection{Connecting to SMBH growth histories in Horizon-AGN}
\label{sec:obs-vs-sims}

As discussed in Sec~\ref{sec:cosmo-sims}, different modes of SMBH growth leave
distinct imprints on their angular momenta. Hence, once integrated over the
lifetime of individual black holes, one would expect different SMBH growth
histories, e.g. accretion- vs. merger-dominated scenarios, to yield
distinguishable predictions for the measurement of their spin parameter \spin.
Indeed, \citealt{Beckmann2023} (hereafter B23) study the existence of such
signatures in the $z=0.0556$ snapshot of the Horizon-AGN cosmological
simulation, finding that black holes grown exclusively via~accretion (nearly
merger-free) have a~spin distribution that is distinct from the rest of the
black hole population at $> 5 \sigma$ significance. As shown in Fig.~3 in B23,
SMBHs which acquired $< 10$\% of their mass in mergers have a~significantly
narrower distribution of spin parameter values than the remainder of the black
hole population, concentrated at spin values close to $a^\ast = 0.998$. The
striking contrast between \spin\ populations reported by the authors presents
a~promising prospect for extending the study into two dimensions - distinguishing 
SMBH growth histories with measurements in the \spinmass\ plane.

To extract predictions for the observable signatures of different SMBH growth
histories in the spin-mass plane from Horizon-AGN, we first identify the loci 
occupied by each history in the \spinmass\ parameter space. 
From here onwards we focus on the $z = 0.0556$ snapshot of the simulation studied in B23, 
which is well suited for our intended target sample of local AGN 
(see Section~\ref{sec:bass-parent}). Inspired by B23, 
we use a~threshold on $f_{\rm BH,\, merge}$, the fraction of mass acquired in
mergers, to identify the \textit{\acc} and \textit{accretion+mergers}
growth channels. We adopt $f_{\rm BH,\, merge} < 0.1$ for \acc\ 
and $f_{\rm BH,\, merge} > 0.1$ for accretion+mergers, which yield populations
of 2137 and 4714 black holes respectively. Since the \acc\ population
in the simulation is limited to $M_{\rm BH} < 10^{8.5}\ \rm M_\odot$, to
showcase its spin predictions across the entire mass range in Fig.~\ref{fig:tongues}, we extend the
dataset over the missing \mbh\ range with random draws from the \acc\ 
\spin\ probability density function (PDF). More specifically, we calculate the 
\acc\ PDF for $M_{\rm BH} > 10^8 {\rm M_\odot}$ and take random draws 
from it to generate \spinmass\ value pairs with \mbh\ distribution matching 
that of the accretion+mergers population beyond $10^8 {\rm M_\odot}$. 
In this way we mimic the effect of switching off spin evolution due to merger
events studied for the $z=0$ snapshot in \cite{Dubois2014b} (see Fig.~7 in the
publication), arriving at an expected \spin\ measurement across the whole range 
in \mbh\ for the \acc\ growth channel. We note that although the
Eddington-limited subgrid model for SMBH in Horizon-AGN does not produce
merger-free populations at very high \mbh, one could still expect highly spinning black holes at 
the high-mass end from models which include super-Eddington accretion rates
in cosmological hydrodynamical simulations \citep[e.g.][]{Massoneau2023}\footnote{We also note that efficient SMBH merging
is, in part, a~consequence of the limitations associated with modelling black 
hole mergers in cosmological simulations, which do not explicitly follow the 
SMBH orbital angular momentum loss via dynamical friction}.

Fig.~\ref{fig:tongues} compares loci in the \spinmass\ plane occupied by black
holes with \acc\ (blue hatched contours) and accretion+mergers (orange
filled contours) growth histories in Horizon-AGN. Current observational
constraints for $M_{\rm BH} > 10^6 {\rm M_\odot}$ are shown as individual
grey points, with open diamonds corresponding to lower limits and filled
diamonds representing measurements and their corresponding $1\sigma$ 
uncertainties. To ensure a~meaningful comparison between observations and the
cosmological simulation, both of the 2D spin distributions extracted from
Horizon-AGN are corrected for the $\eta(a^\ast)-a^\ast$ bias and hence are
vertically shifted upwards with respect to the raw output of the simulation. Fig.~\ref{fig:tongues} clearly demonstrates that the \textit{two SMBH growth 
scenarios yield different signatures in the spin--mass
parameter space}. At $M_{\rm BH}
\gtrsim 10^8 {\rm  M_\odot}$ the two growth histories begin separating, with
mergers pushing the expected spin parameter values down towards values as
low as $a^\ast \sim 0.6$ with increasing \mbh, while the \acc\ scenario remains
concentrated at near-maximal spin values. For black hole masses $<10^8\ M_\odot$ both \acc\ and
accretion+mergers populations are concentrated around near-maximal \spin\
values. This degree of overlap is further exacerbated by the spin-dependent
radiative efficiency correction, which shifts the orange contours upwards,
bringing the two populations towards near-identical loci. Overall, in the context of 
the Horizon-AGN cosmological simulation, Fig.~\ref{fig:tongues} presents a~promising opportunity
for differentiating between SMBH growth channels in X-ray observations of AGN,
even in the presence of the expected $\eta(a^\ast)-a^\ast$ bias.

\section{The need for a~systematic study of the spin-mass plane}
\label{sec:need-for-study}

Another critical conclusion drawn from Fig.~\ref{fig:tongues} is the
necessity of obtaining high-precision spin measurements for large 
statistical samples of local AGN.
Although the current observational constraints are broadly consistent with
the accretion+mergers growth channel and show a~hint of decrease in \spin\
with \mbh\ above $10^8 \rm M_\odot$, the limited number of measurements at
these masses and the relatively poor precision of these measurements mean
the current data~are not sufficient for a~statistical assessment of these
trends. More importantly for our discussion, the current sample of 10
measurements and 23 upper limits presented in the figure is not capable of
differentiating between the two signatures of SMBH growth histories
predicted by Horizon-AGN.

The current state-of-the-art spin measurements also offer a~rather limited
opportunity for the validation and improvement of the SMBH spin evolution
models used in cosmological simulations of SMBH growth. When compared
against other hydrodynamical simulations which cover a~similar range in
\mbh\ \citep{Bustamante19}, Horizon-AGN produces spin population statistics
with only subtle differences in their \spinmass\ loci. In general this is
encouraging, as it suggests that the qualitative trends implied by these
simulations (i.e. Figure \ref{fig:tongues}) are robust. The discrepancies
that do exist between them, however, are a~combined result of differences in
subgrid prescriptions for SMBH seeding, accretion and feedback, and hence
carry important information about the validity of a~given numerical approach.
With the measurement precision and sample size offered by the current
constraints, one cannot determine which model implementation (if any) is a
closer approximation of the observable Universe. Consequently, the current
constraints allow enough room to validate a~range of models, while, at the
same time, are not strong enough to serve as observational calibrators
for future generations of subgrid cosmological SMBH models.

In summary, Fig.~\ref{fig:tongues} demonstrates that {\bf{\it a~systematic
study of the spin-mass plane is necessary for differentiating between
different growth histories of SMBHs in the local Universe}}. A~comprehensive
characterisation of \mbh\ and \spin\ properties of local AGN will also play
a~critical role in calibrating subgrid models of cosmological structure
formation by delivering improved constraints on physical properties of SMBH.
To establish measurement requirements for these survey observations we take
the two SMBH growth histories in Horizon-AGN as a~case study, and investigate
the precision and sample sizes that would be required to differentiate
between these two possibilities in realistic samples of observable sources.

\section{The BASS Sample} 
\label{sec:bass}

In order to assess the number of known AGN for which spin constraints may be possible,
either now or in the moderately near future, we turn to the BAT AGN Spectroscopic
Survey (BASS; \citealt{BASS}). This is a~major multi-wavelength effort to characterise
AGN detected in the very high energy X-ray survey conducted by the Burst Alert
Telescope (BAT, 14--195\,keV; \citealt{SWIFT_BAT}) onboard the \textit{Neil Gehrels
Swift Observatory} (hereafter \textit{Swift}; \citealt{SWIFT}). The latest 105-month
release of the BAT source catalogue (\citealt{BAT105m}) contains 1632 sources in
total, of which 1105 have been identified as some kind of AGN. BASS provides black
hole mass estimates (drawn from a~variety of methods including H$\beta$ line widths,
stellar velocity dispersions and literature searches\footnote{The method used to
estimate the mass for each source is also provided in the BASS catalogue.}) and
Eddington ratios for the majority of these (e.g. \citealt{BASSDRII}), as well as
initial constraints on their broadband spectral properties by combining the BAT data
with the best soft X-ray coverage available (e.g. from \textit{XMM-Newton}; 
\citealt{Ricci17}).

\subsection{Sources with spin measurement potential}
\label{sec:bass-parent}

Not all AGN are well suited to making spin measurements. In particular, the very hard
X-ray selection of the BAT survey means it is sensitive to even heavily obscured AGN
(including `Compton thick' sources with $N_{\rm{H}} > 1.5 \times 10^{24}$ cm$^{-2}$;
e.g. \citealt{Arevalo14, Annuar15}), and determining the contribution from
relativistic reflection becomes increasingly challenging as the source becomes more
obscured. Furthermore, there is a~general expectation that at low accretion rates the
optically-thick accretion disk becomes truncated at radii larger than the ISCO
(\citealt{Narayan94, Esin97, Tomsick09}), such that the inner radius of the disk no
longer provides direct information about the spin. As such, in order to compile 
a~sample of AGN for which spin measurements should be possible, we apply several 
selection criteria to the BASS sample. First, we select sources with fairly low levels of obscuration,
requiring a~neutral column density of $N_{\rm{H}} \leq 10^{22}$ cm$^{-2}$. Second, we
select sources with Eddington ratios of $\lambda~= L_{\rm{bol}}/L_{\rm{Edd}} \geq
0.01$, so that we may have confidence that the accretion disk should extend to the
ISCO (note also that this selection naturally means a~black hole mass estimate is
available, otherwise it would not have been possible to estimate $\lambda$). 
We further place an upper limit of $\lambda < 0.7$ to select sources for which
thin disk approximation is appropriate \citep[e.g.][]{Steiner2010}. We also
require that the sources exhibit sufficiently strong reflection features in order for
spin measurements to be feasible. We therefore then select sources for which the
initial X-ray spectroscopy conducted by \cite{Ricci17} indicates a~reflection fraction
consistent with $R \geq 1$ (note that \citealt{Ricci17} use the definition of the
reflection fraction from \citealt{pexrav}). For the same reason, we also exclude
blazars from our sample. Finally, after having made these cuts to the BASS data, we
also exclude a~small number of remaining sources for which the initial X-ray
spectroscopy implies an unreasonably hard photon index ($\Gamma~\leq 1.5$), taking this
as an indication that either the source is actually obscured but that the absorption is
sufficiently complex that it was not well characterised by the simple spectral models
used in \cite{Ricci17} or that the spectrum has a low signal-to-noise ratio.

We do not expect these selections to introduce any significant biases that would
cause the observed spin distribution to deviate from the intrinsic one. Selecting
based on obscuration properties and the exclusion of blazars are both expected to be
related mainly to our viewing angle to these AGN (e.g. \citealt{Antonucci1993}),
which is not dependent on the spin of their central SMBHs. The Eddington ratio
selection relates specifically to the accretion rate onto the AGN `today', while
the spin of the SMBH will be mainly determined by its long-term growth history
instead. Indeed, the Horizon-AGN simulation shows no connection between SMBH spin and the instantaneous accretion rate in the analysed snapshot. Finally, the reflection fraction selection
still permits spins across the full range of possible prograde spin values ($a^* = 0 - 0.998$), as even Schwarzschild black holes
($a^*$ = 0) should result in reflection from the disc with $R \geq 1$ if the disc
reaches the ISCO \citep[e.g.][]{Dauser2016}.

\begin{figure}[h]
    \centering
    \includegraphics[width=\textwidth]{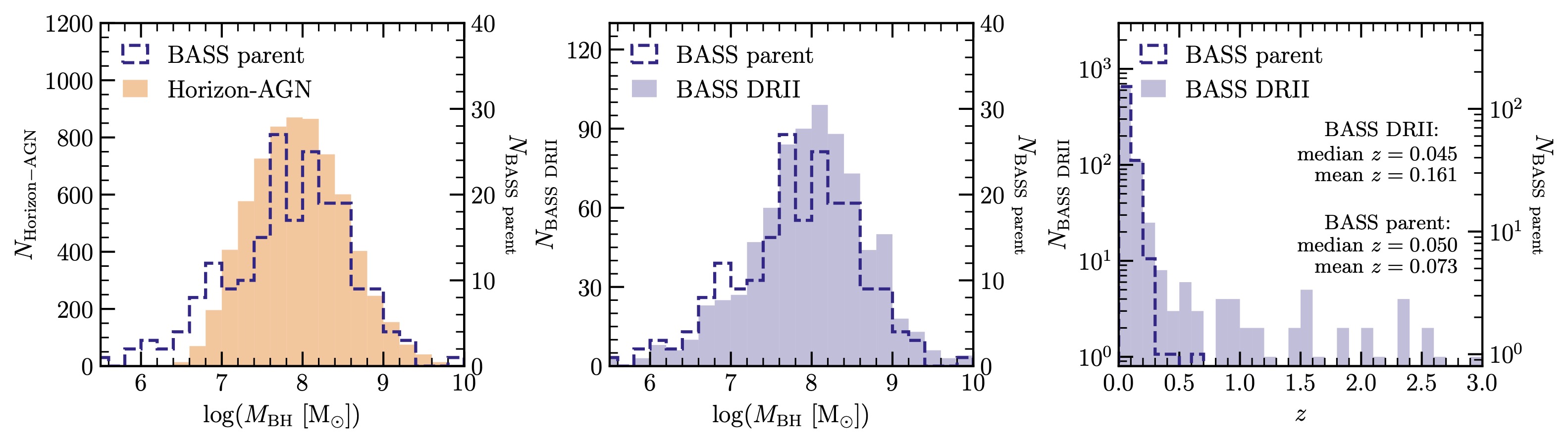}
    \caption{Left panel: comparison of SMBH mass distributions
    between our parent BASS sample (BASS parent, Section~\ref{sec:bass-parent} 
    and the Horizon-AGN 
    cosmological simulation, demonstrating good agreement
    between potential AGN targets and the cosmological model.
    Middle panel: \mbh\ comparison between the whole BASS catalogue 
    (BASS DRII) and out parent BASS sample. The distributions in $M_{\rm BH}$ are comparable, 
    demonstrating that our sample selection criteria~do not 
    introduce additional bias with respect to the full BASS DRII catalogue. 
    Right panel: comparison of redshift distributions between our 
    parent BASS sample and the full BASS DRII AGN catalogue sources. 
    As expected, our sample selection removes highest-$z$ sources, 
    however the median $z$ remains comparable at $z\sim 0.05$.}
    \label{fig:bass-sample}
\end{figure}

Our various cuts leave a~sample of 192 AGN from BASS for which spin measurements
should be possible. All of these sources are relatively local, with a~median
redshift of $z \sim 0.05$ (see 'BASS parent' in the right panel of Figure \ref{fig:bass-sample}), consistent with 
the $z=0.0556$ snapshot of the Horizon-AGN simulation we use in this work. Conveniently, 
these sources span a~broad range of black hole masses, 
$5.5 \lesssim \log(M_{\rm{BH}} / \rm M_\odot) \lesssim 10.0$, (middle and left panels in Fig.~\ref{fig:bass-sample}) which is well suited to
placing constraints on different potential black hole growth models. The sample is
also particularly appropriate for comparisons against Horizon-AGN specifically,
with the \mbh\ distribution in our parent BASS sample showing good qualitative agreement with the
black holes extracted from the cosmological simulation (left panel in Fig.~\ref{fig:bass-sample}).

\section{Sample size and spin uncertainty requirements}
\label{sec:sample-reqs}

In order to design observational strategies for differentiating between supermassive
black hole growth scenarios with spin parameter measurements, we need to establish 
the sample size and 
maximum spin measurement uncertainty (\dspin) necessary for such work. Both requirements are
imposed by a~combination of two independent factors: the difference between the SMBH
growth history models as a~function of mass in Horizon-AGN, and the SMBH mass
distribution of our parent BASS sample (Section \ref{sec:bass-parent}). In order
to determine these requirements, below we simulate a~set of potential survey
programmes covering a~range of realistic measurement uncertainties and sample sizes.

\subsection{Simulating Realistic Samples Informed by Horizon-AGN}
\label{sec:sim-setup}

Fig.~\ref{fig:tongues} shows that the two SMBH growth scenarios separate in the 
\spinmass\ plane above $10^8\, \rm M_\odot$, with the strongest difference at the
high-\mbh\ end. It is also apparent that at high masses the \accmer\ 
distribution (orange) spans a~broad range in spin parameter values. Therefore, in
order to differentiate between the two growth scenarios with realistic samples, we
base our strategy on a simple statistical approach - discretising the spin--mass 
plane by calculating sample means in bins of \mbh.
This way we can both reduce the impact of the uncertainty on individual \spinmass\
measurements and coarsely capture the two-dimensional trends which would
otherwise require large sample sizes for a~direct comparison between 2D
distributions. To best capture the signal we select one bin below $M_{\rm BH} =
10^8 ~\rm M_\odot$ (where both scenarios make consistent predictions) and for higher
masses we split the sample into two bins above and below $\log(M_{\rm BH}/ M_\odot)
= 8.7$ ($M_{\rm BH} \approx 5\times 10^8 ~\rm M_\odot$). This choice is motivated by
the SMBH mass distribution shown in panel A~of~Fig.~\ref{fig:bass-sample} - there
are 9 AGN with $\log(M_{\rm BH}/ M_\odot) > 8.7$ among the brightest 100
X-ray sources in the 2-10 keV band in our parent BASS selection 
(which we consider as an
initial plausible sample size), which yield a~Poisson signal-to-noise ratio of 3 for source number count in the highest mass bin. On the low-mass end we restrict the mass bin to $M_{\rm BH} > 10^6\ \rm M_\odot$ to match Horizon-AGN \mbh\ distribution.
Our final binning scheme results in the following ranges in $\log(M_{\rm BH} /
\rm M_\odot)$: [6.0, 8.0], [8.0, 8.7] and [8.7, 10.0]. 

By estimating mean \spin\ values, we take advantage of the $\sqrt{M}$ scaling with 
sample size $M$ of the standard error on the mean, which allows us to clearly
separate the two SMBH growth scenarios in the coarsely binned \spinmass\ plane.
With \mbh\ ranges in place, we can now determine the minimum sample size and uncertainty
required for a~series of mean \spin\ measurements to differentiate between the
orange and blue SMBH populations in Fig.~\ref{fig:tongues}. We proceed by selecting
the $K$ brightest sources from our parent BASS sample for a~set of different values
for $K$, thus imposing a~single X-ray flux limit across the whole range in SMBH mass
in each case. This way we match the assumptions discussed earlier in
Sec.~\ref{sec:current-constraints} and hence can implement the radiative efficiency
- spin parameter bias expected for flux-limited AGN samples in the local Universe.
We then extract \spin\ distributions for both SMBH growth histories in each \mbh\ 
bin from Horizon-AGN and correct them for the \radeff\ bias (see Appendix for
details). 

To simulate realistic samples of AGN, we take a~total of $ K = \sum K_i$ random 
draws from the bias-corrected spin distributions across all $M_{\rm BH}$ bins, with the sample 
sizes $K_i$ in each individual bin determined by the number of BASS sources 
which fall within this SMBH mass range\footnote{Throughout the study we treat 
the 3 sources with $M_{\rm BH} < 10^6\ \rm M_\odot$ in the parent BASS sample 
as belonging to the lowest mass bin, given the large uncertainty associated
with low \mbh\ estimates}, among the $K$ brightest AGN from our selection
in Sec.~\ref{sec:bass-parent}.
We perform the spin draws separately for each SMBH growth
model, and hence for each realisation of a~potential observational sample of 
a total $K$ AGN
we have a~paired set of predicted spin values for the \acc\ and \accmer\ 
growth models. To simulate the effect of measurement uncertainty we then
perturb the spin values drawn above by further random draws from a~normal
distribution $N(0, \sigma_{a^\ast})$, where $\sigma_{a^\ast}$ is the assumed
uncertainty on the spin measurements\footnote{Note that at this point we simply
assume all spin measurements to have the same uncertainty; more realistic scenarios
that include the expected dependence of measurement uncertainties as a
function of spin are considered below in Section \ref{sec:job-done}.}, for a~range
of values for $\sigma_{a^\ast}$. For each $a^\ast_i$, the normal distribution
is singly truncated at $\sigma_i = (0.998 - a^\ast_i)/\sigma_{a^\ast}$ to account for
the theoretical upper limit on the spin parameter of $a^\ast = 0.998$. 
This way, each potential AGN sample from a~given
SMBH growth history consists of $K$ spin measurements expected from \radeff\
bias-corrected Horizon-AGN distributions, post-processed to account for measurement
uncertainty.
 
As a~final step in each simulated AGN sample, for each mass bin we calculate the
mean spin value together with its corresponding standard error on the mean:
$\mu_{\rm acc} \pm \delta_{\rm acc}$ for the \acc\ scenario and 
$\mu_{\rm acc+mer} \pm \delta_{\rm acc+mer}$ for \accmer.
To calculate the standard error, we necessarily assume that each paired draw
consists of \spin\ measurements without lower limits in both SMBH growth
histories, however, we note that some are likely to be present in real observations. 
Finally, we check the probability of the two mean measurements 
in a~mass bin $\Omega$ being drawn from the same underlying distribution 
by calculating the right-tail p-value ($p_{\Omega}$) of $\mu_{\rm acc} -
\mu_{\rm acc+mer}$ belonging to $N(0, \sqrt{\delta_{\rm acc}^2 +
\delta_{\rm acc+mer}^2})$. This way, we calculate the probability that 
a~given paired set of predicted spin values in a~bin cannot differentiate
between the \acc\ and \accmer\ growth scenarios.
For our simulated AGN samples we consider spin measurement uncertainties 
in the range $\sigma_{a^\ast} \in [0.02, 0.05, 0.10, 0.15, 0.20]$, and sample 
sizes of $K \in [50, 70, 100, 150, 192]$. For each combination of these values we repeat
the above process $10^4$ times, generating a~large set of potential AGN samples,
together with their corresponding mean spin parameter measurements as a~function of
mass.

\begin{figure}
    \centering
    \includegraphics[width=0.5\textwidth]{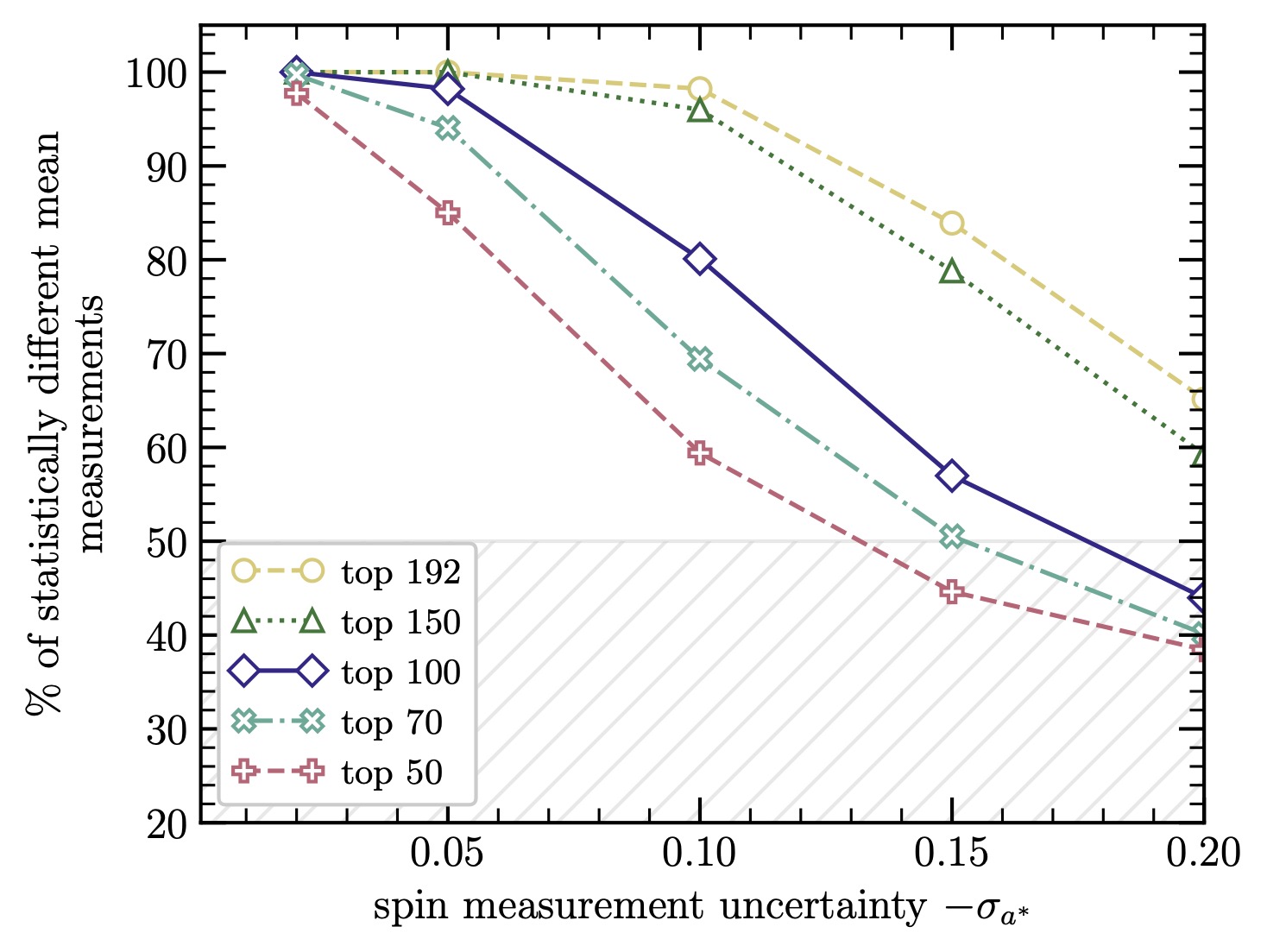}
    \caption{Minimum spin measurement uncertainty ($\sigma_{a^\ast}$) and sample size 
    required for simulated observations to differentiate between \acc\ and 
    merger \& accretion SMBH growth scenarios drawn from the HorizonAGN cosmological 
    simulation. Individual curves show the simulated observations 
    performed for the top 50, 70, 100, 150 and 192 brightest X-ray sources 
    in the 2-10 keV band in our parent BASS sample. In order for a~median 
    (50th percentile) simulated measurement to 
    tell SMBH growth histories apart, one requires Gaussian spin measurement uncertainty 
    of $\lesssim 0.15$ across the entire range in $a^\ast$ for a~sample of 70 sources.}
    \label{fig:sigma-target}
\end{figure}

Fig.~\ref{fig:sigma-target} shows the fraction of these simulated AGN samples
which result in a~statistically different measurement of mean spin parameter
value between the \acc\ and accretion+mergers SMBH growth scenarios as
a~function of \spin\ measurement uncertainty and sample size. In order to take
full advantage of the difference between the two predictions at high SMBH
masses, in each realisation of a~potential AGN sample we combine the results
for both of the mass bins with $M_{\rm BH} > 10^8 \rm M_\odot$. Since these
can be treated as independent measurements, the combined probability of the
simulated \acc\ and accretion+mergers samples being drawn from the
same distribution in the \spinmass\ plane for \textit{both} mass bins becomes
$p_{\rm total} = p_{\rm [8,8.7]} \times p_{\rm [8.7,10]}$. The vertical axis
in Fig.~\ref{fig:sigma-target} shows the fraction of simulated measurements
which result in $p_{\rm total} < 0.01 $ as a~function of $\sigma_{a^\ast}$ in
the x-axis. Different lines and markers correspond to the top 192, 150, 100, 70 
and 50 brightest AGN in our parent BASS sample. Whenever the
curves enter the gray hatched region, the median measurement expected from the
$10^4$ realisations of potential AGN samples will not differentiate between
the \acc\ and \accmer\ BH growth scenarios.

As expected, Fig.~\ref{fig:sigma-target} shows that increasing the uncertainty 
on individual spin measurements at 
fixed sample size decreases the fraction 
of simulated samples capable of differentiating between the two SMBH growth
histories. Similarly, at fixed uncertainty, a~larger sample size results in
higher chances for a~single realisation of such a~sample to tell the two growth
histories apart. In the context of minimum survey requirements, we find that for~a
sample size of $>$70 AGN we would need a~maximum of $\sigma_{a^\ast} \lesssim
0.15$ for a~median expected measurement to differentiate between the
\acc\ and \accmer\ SMBH populations. In order to be
conservative, we therefore suggest that a~sample size of $N = 100$ AGN drawn
from our BASS selection with spins measured to a~precision of $\sigma_{a^\ast}
\sim 0.15$ (or better) would be sufficient to distinguish between these
different SMBH growth scenarios.

\section{HEX-P Simulations}

Given the importance of simultaneous broadband X-ray spectroscopy for providing
observational constraints on BH spin, a~mission with the profile of \textit{HEX-P}
is ideally suited to building the significant samples of SMBH
spin measurements required to finally connect these measurements to cosmological
BH growth models. We now present a~set of simulations that showcase the
anticipated capabilities of \textit{HEX-P} in this regard.

\subsection{Mission Design}

The low-energy telescope (LET) onboard \hexp\ consists of a~segmented mirror assembly coated with Ir on monocrystalline silicon that achieves a~half power diameter of $3.5''$, and a~low-energy DEPFET detector, of the same type as the Wide Field Imager (WFI; \citealt{wfi}) onboard Athena~\citep{Nandra13}. It has $512 \times 512$ pixels that cover a~field of view of $11.3'\times 11.3'$. It has an effective passband of 0.2--25\,keV, and a~full frame readout time of 2\,ms, which can be operated in a~128 and 64 channel window mode for higher count-rates to mitigate pile-up and faster readout. Pile-up effects remain below an acceptable limit of ${\sim}1\%$ for fluxes up to ${\sim}100$\,mCrab in the smallest window configuration (64w). Excising the core of the PSF, a~common practice in X-ray astronomy, will allow for observations of brighter sources, with a~typical loss of up to ${\sim}60\%$ of the total photon counts.

The HET consists of two co-aligned telescopes and detector modules. The optics are made of Ni-electroformed full shell mirror substrates, leveraging the heritage of XMM-Newton, and coated with Pt/C and W/Si multilayers for an effective passband of 2--80\,keV. The high-energy detectors are of the same type as flown on \nustar, and they consist of 16 CZT sensors per focal plane, tiled $4\times4$, for a~total of $128 \times 128$ pixel spanning a~field of view of $13.4' \times 13.4'$.

\subsection{Instrumental Responses}

All the mission simulations presented here were produced with a~set of response files that represent the observatory performance based on current best estimates (see \textbf{[Madsen et al.\ 2023]}). The effective area~is derived from raytracing calculations for the mirror design including obscuration by all known structures. The detector responses are based on simulations performed by the respective hardware groups, with an optical blocking filter for the LET and a~Be window and thermal insulation for the HET. The LET background was derived from a~GEANT4 simulation \citep{Eraerds2021} of the WFI instrument, and the one for the HET from a~GEANT4 simulation of the \nustar\ instrument, both positioned at L1. The broad X-ray passband and superior sensitivity will provide a~unique opportunity to study accretion onto SMBHs across a~wide range of energies, luminosities, and dynamical regimes.

\subsection{Spectral Simulations} 
\label{sec:spectral-sims}

To assess the ability of \textit{HEX-P} to constrain black hole growth models via
AGN spin measurements, we perform a~series of spectral simulations covering a~range
of spin values relevant to the cosmological simulations discussed above: $a^* = 0.5,
0.7, 0.9, 0.95$. We first focus on the lowest spin value among this set, $a^* =
0.5$, and perform a~set of simulations at different exposures in the range
50--200\,ks. For each exposure, we simulate 100 different spectra using the above
response files and the \textsc{xspec} spectral fitting package (v12.11.1;
\citealt{XSPEC}), in order to determine the minumum exposure that provies the
\textit{HEX-P} data~quality needed for an average 1$\sigma$ constraint on the spin
of $\sigma_{a^*} \leq 0.15$, as required above. We then perform further sets of 100
simulations with this exposure for each of the spin values listed above to map out
the expected uncertainty as a~function of input spin (as it is now well-established
that for a~given observational setup tighter constraints can be obtained for higher
spin values, e.g. \citealt{Bonson16, Choudhury17, Kammoun18, Barret19}).

We make use of the \textsc{relxill} family of disk reflection models
(\citealt{relxill, Dauser14}) for these simulations, and construct a~spectral model
that draws on the typical properties of the $\sim$200 BASS AGN selected above
(Section \ref{sec:bass-parent}), general expectations for the unobscured AGN selected based
on the unified model for AGN, and typical results seen from AGN for which detailed
reflection studies have already been possible in the literature. We specifically
make use of the \textsc{relxilllpcp} model, which assumes a~simple lamppost geometry
for the corona, and that the ionising continuum is a~thermal Comptonization model
(specifically the \textsc{nthcomp} model; \citealt{nthcomp1, nthcomp2}). 

For the key model parameters, we assume the spectrum of the primary continuum has
parameters $\Gamma~= 2.0$ and $kT_{e} = 100$\,keV for the photon index and electron
temperature, respectively, relatively typical parameters for unobscured AGN
(\citealt{Ricci17, Balokovic20}). This illuminates a~geometrically thin accretion
disk which is assumed to extend down to the ISCO. AGN with moderate-to-low levels
of obscuration are generally expected to be viewed at moderate-to-low inclinations
in the unified model (\citealt{Antonucci1993}), so we assume an inclination of
$45^{\circ}$. The disk is assumed to be ionised, with an ionisation
parameter\footnote{The ionisation parameter has its usual definition of $\xi =
L/nR^{2}$, where $L$ is the ionising luminosity, $n$ is the number density of the
plasma, and $R$ is the distance between the plasma~and the ionising source.} of
$\log[\xi/(\rm{erg}~\rm{cm}~\rm{s})] = 2$ (typical of values reported from
relativistic reflection analyses in the literature, e.g. \citealt{Ballantyne11,
Walton13spin, Jiang19agn, Mallick22}), and to have solar iron abundance of (i.e.
$A_{\rm{Fe}} = 1$). We assume the corona~has a height of 5\,$R_{\rm{G}}$ for the
$a^* = 0.5$ simulations, and then that it has~constant height in vertical horizon
units as we subsequently vary the spin. This is equivalent to assuming the scale
of the corona~contracts slightly as the inner radius of the disk moves inwards,
as might be expected should the corona~be related to magnetic re-connection
around the inner disk (e.g. \citealt{Merloni01}); in gravitational radii these
coronal heights correspond to values in the range $\sim$4--5\,$R_{\rm{G}}$,
which are relatively standard values inferred for the sizes of X-ray coronae
(e.g. \citealt{Reis13corona, deMarco13, Kara16, Mallick2021}). The reflection fraction is
calculated self-consistently based on the combination of $a^*$ and $h$, both
when computing the simulations and also when subsequently fitting the simulated
data~to explore the constraints on $a^*$.

\begin{figure}[h!]
    \centering
    \includegraphics[width=0.5\textwidth]{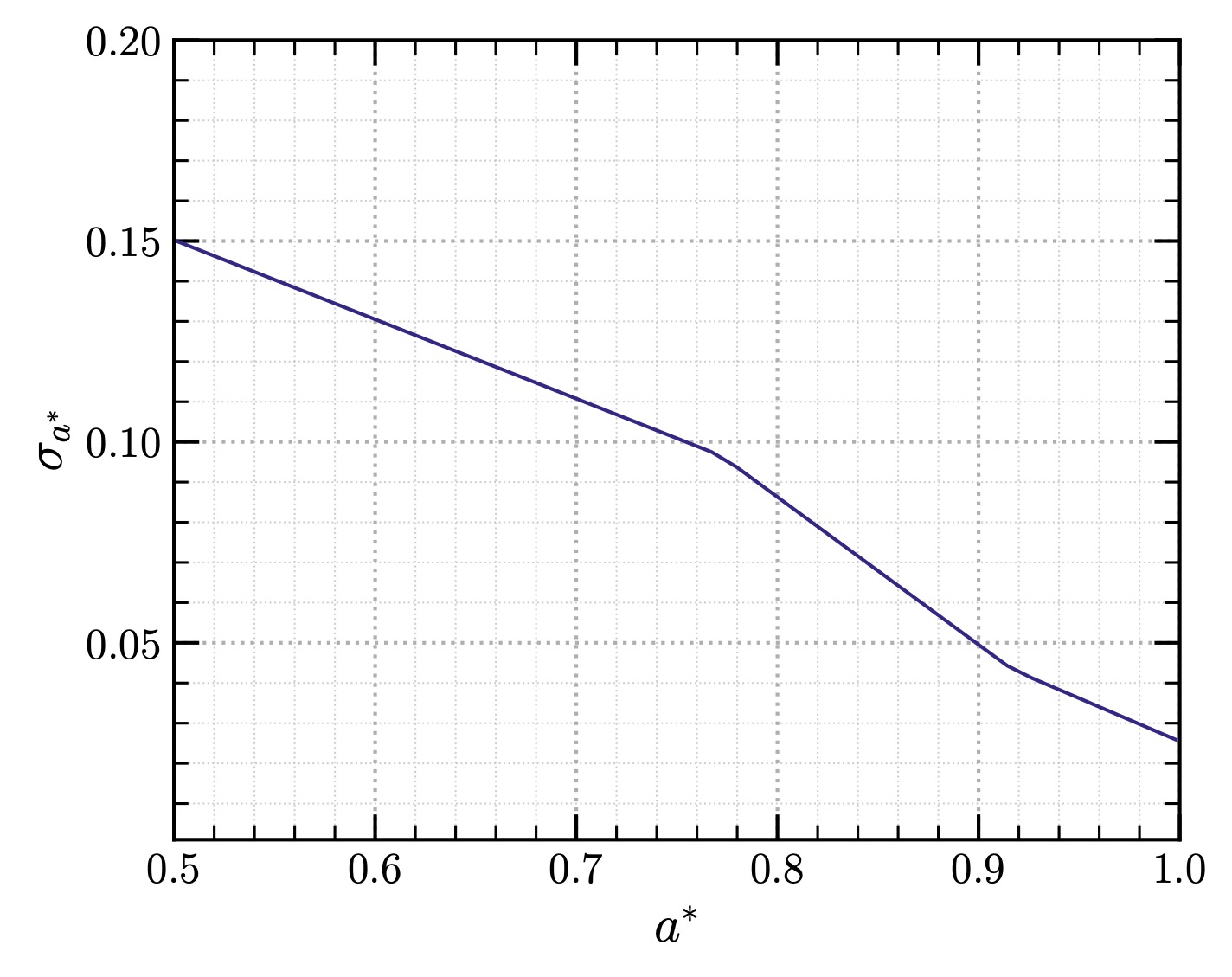}
    \caption{Spin parameter measurement uncertainty $\sigma(a^\ast)$ 
    vs spin parameter $a^\ast$, extracted from the \textit{HEX-P} spectral simulations of relativistic reflection in AGN discussed in Section \ref{sec:spectral-sims} (which assume a~2--10\,keV flux of $6 \times
    10^{-12}$\,erg~cm$^{-2}$~s$^{-1}$ and an exposure time of 150\,ks).}
    \label{fig:hexp-sigma}
\end{figure}

\begin{figure*}
\begin{center}
\hspace*{-0.4cm}
\rotatebox{0}{
{\includegraphics[width=235pt]{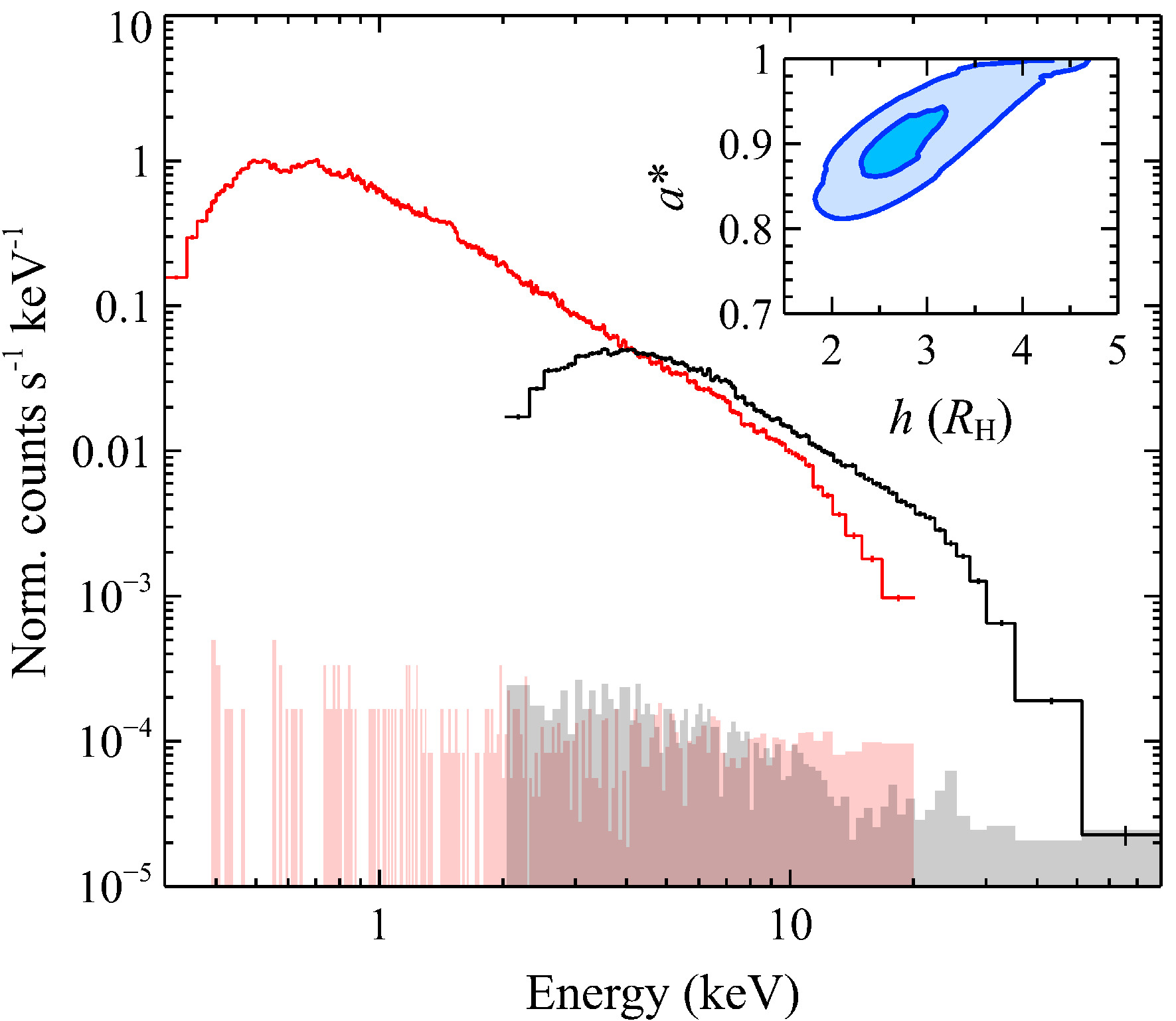}}
}
\hspace*{0.4cm}
\rotatebox{0}{
{\includegraphics[width=235pt]{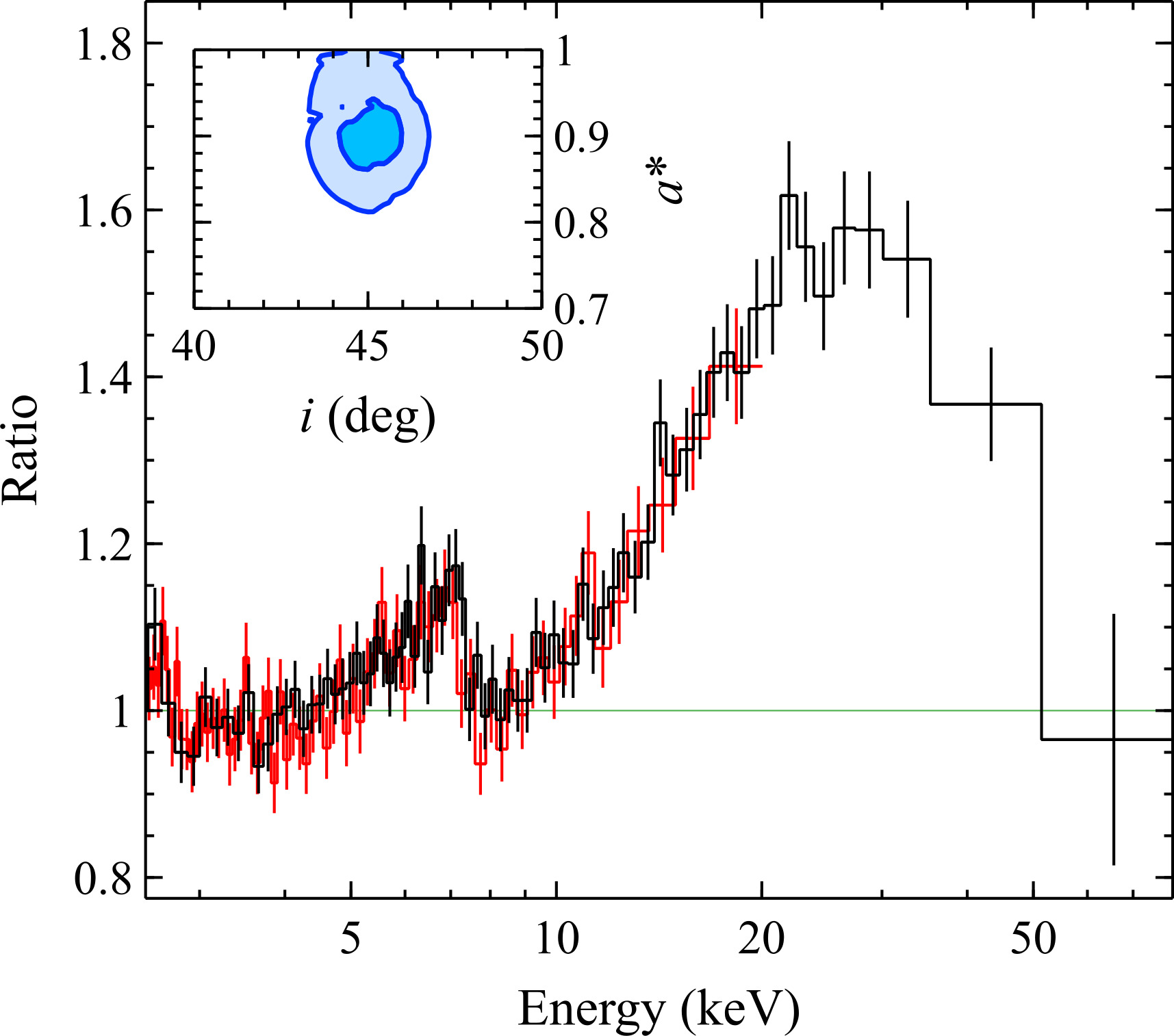}}
}
\end{center}
\vspace*{-0.3cm}
\caption{
\textit{Left panel:} Count spectra~for the LET (red) and the combined HET (black)
for one of our \textit{HEX-P} spectral simulations with $a^* = 0.9$, along with
their corresponding instrumental backgrounds (shaded regions).
\textit{Right panel:} The same simulated spectra plotted as a ratio to a simple
\textsc{cutoffpl} continuum model (a powerlaw with an exponential high-energy
cutoff), fit to the 2--4, 8--10 and 50--80\,keV bands (where the primary continuum
would be expected to dominate). We zoom in on the data above 2 keV to highlight
the key reflection features: the relativistically broadened Fe K$\alpha$ peaking
at $\sim$6--7\,keV and the Compton hump peaking at $\sim$30\,keV. The two insets
show the projected constraints on $a^*$, $h$ and $i$ for this simulation
(displayed as 2-D confidence contours, with the 1$\sigma$ and 2$\sigma$ levels
shown).}
\label{fig:simspec}
\end{figure*}

As we need a~sample of $\sim$100 AGN to distinguish the black hole growth models,
we normalise all of our simulations to have a~2--10\,keV flux of $6 \times
10^{-12}$\,erg~cm$^{-2}$~s$^{-1}$, as this is the flux of the 100th brightest
source in the sample of AGN selected above (see Section \ref{sec:bass-parent}). At this
flux level, we find that a~\textit{HEX-P} exposure of 150\,ks is required to
provide an average uncertainty of $\sigma_{a^*} \leq 0.15$ for an input spin value
of $a^* = 0.5$. We also find that, as broadly expected, the typical uncertainty
on the spin measurements \textit{HEX-P} will decrease for more rapidly
rotating black holes (see Figure \ref{fig:hexp-sigma}). Although our results do
differ quantitatively to the prior efforts that have also come to similar
conclusions, the overall quantitative trend we find is consistent with these
previous works (e.g. \citealt{Bonson16, Kammoun18}). These quantitative differences
mainly relate to the fact that we are simulating \textit{HEX-P} spectra~instead
of \textit{XMM-Newton}+\textit{NuSTAR} spectra, with some contribution from the
fact that different model versions and slightly different assumptions are used
here. An example spectrum from one of our simulations for $a^* = 0.9$ is shown in
Figure \ref{fig:simspec}, along with the projected constraints on $a^*$, $i$ and
$h$ for that simulation.

\section{Constraining SMBH growth channels with \textit{HEX-P}}
\label{sec:job-done}

Having established both the minimum survey requirements and the \hexp\ S/N that will
be required to meet these, we now demonstrate the mission's ability to differentiate 
between \acc\ and \accmer\ supermassive black hole growth scenarios. To this end we
repeat the simulations introduced earlier in Sec.~\ref{sec:sim-setup} for the
brightest 100 sources in our parent BASS sample (Sec.~\ref{sec:bass-parent}), this
time replacing the constant spin parameter uncertainty \dspin\ assumed previously
with the $a^\ast$-dependent uncertainties $\sigma(a^\ast)$ obtained from spectral
simulations in Sec.~\ref{sec:spectral-sims} (see Fig.~\ref{fig:hexp-sigma}), i.e.
now specifically simulating observational campaigns with \textit{HEX-P}. Similarly
to our initial tests, we simulate $10^4$ paired AGN samples to asses the probability
with which \hexp\ will confidently discriminate between SMBH growth histories.

\begin{figure}
    \centering
    \includegraphics[width=\textwidth]{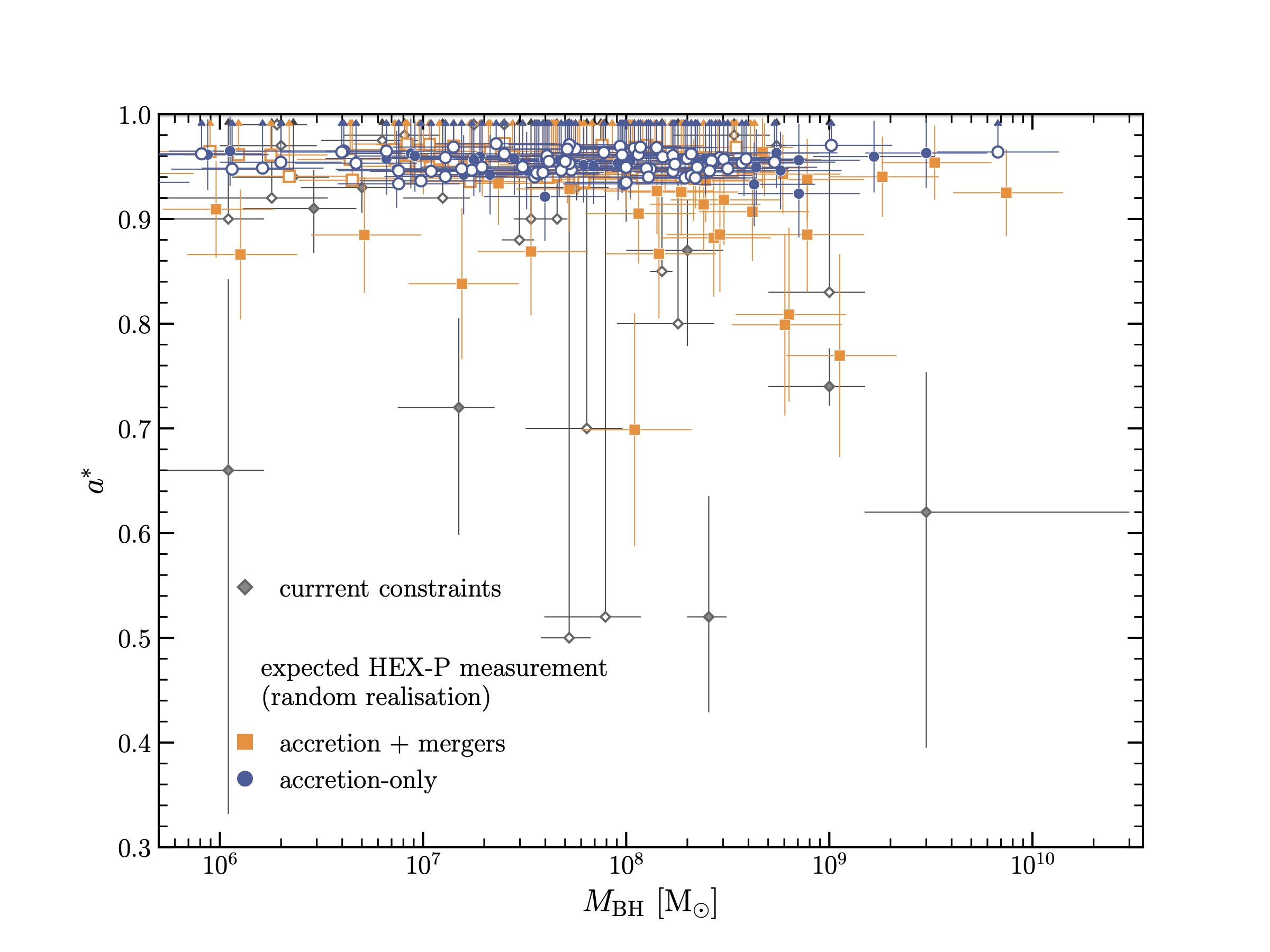}
    \caption{Comparison between current published constraints (grey diamonds) 
    and a~random realisation of simulated HEX-P measurements for the \acc
    (blue circles) and \accmer\ SMBH growth scenarios (orange squares). 
    Empty symbols show lower limits on \spin, while filled symbols denote measurements
    with vertical errorbars marking their corresponding $1\sigma$ confidence ranges. 
    With 
    its expected spin measurement uncertainty \textit{HEX-P will provide unprecedented
    constraints on the \spinmass\ plane for SMBH in the local Universe.}}
    \label{fig:example-measurement}
\end{figure}

Fig.~\ref{fig:example-measurement} illustrates the observational constraints that
would be expected for one such paired set of simulated AGN samples, randomly selected
from our set of such simulations, comparing the \acc\ (blue circles) and \accmer\
(orange squares) SMBH growth scenarios against the current spin parameter constraints
for \mbox{$M_{\rm BH} > 10^6 \rm M_\odot$} (grey diamonds). In order to account for
the existence of lower limits expected in realistic measurements, by analogy with the
measurements collected from the literature, we use open symbols to indicate
right-censored spin draws, i.e. $a^\ast_i$ for which \mbox{$0.998-a^\ast_i <
\sigma(a^\ast_i)$}. Vertical error bars mark $1\sigma$ uncertainties in both simulated 
and existing \spinmass\ constraints, allowing 
for a~direct comparison between the two. All
BASS \mbh\ measurements in the figure are assigned a $\pm \rm 0.3~dex$ uncertainty
expected for their estimate via the $M_{\rm BH} - \sigma_\ast$ relation
\citep{KormendyHo2013}. 
Fig.~\ref{fig:example-measurement} clearly demonstrates the
potential for such a \hexp\ spin measurement campaign to map the spin-mass plane with 
unprecedented accuracy. In comparison with currently available constraints, 
the \textit{simulated observations offer visible improvements on measurement precision,
good control of sample selection biases and an over twofold increase in sample size}.

\begin{table}
    \caption{Summary of mean spin parameter values in simulated \hexp\ observations
             and Horizon-AGN}
    \centering
    
    \begin{tabular}{llccc}
     \toprule
     $\log(M_{\rm BH}/ \rm M_\odot )$ bin & & [6.0, 8.0] & [8.0, 8.7] & [8.7, 10.0]  \\
     \midrule
     \multirow{2}{*}{Horizon-AGN} & \acc & $0.996$ & $0.993$ & $0.993$ \\
     & \accmer & $0.959$ & $0.905$ & $0.768$ \\
     \\[2pt]
     Horizon-AGN  & \acc & $0.996$ & $0.995$ & $0.995$ \\
     with \radeff\ bias & \accmer & $0.981$ & $0.953$ & $0.851$ \\
     \\[2pt]
     \multirow{2}{*}{Expected measurement} & \acc & $0.977\pm 0.002$ & $0.975 \pm 0.003$ & $0.976 \pm 0.005$ \\
      & \accmer & $0.966 \pm 0.004$ & $0.945 \pm 0.011$ & $0.86 \pm 0.04$ \\
      \bottomrule
    \end{tabular}
    \label{tab:mean-stats}
\end{table}

\begin{figure}[h!]
    \centering
    \includegraphics[width=\textwidth]{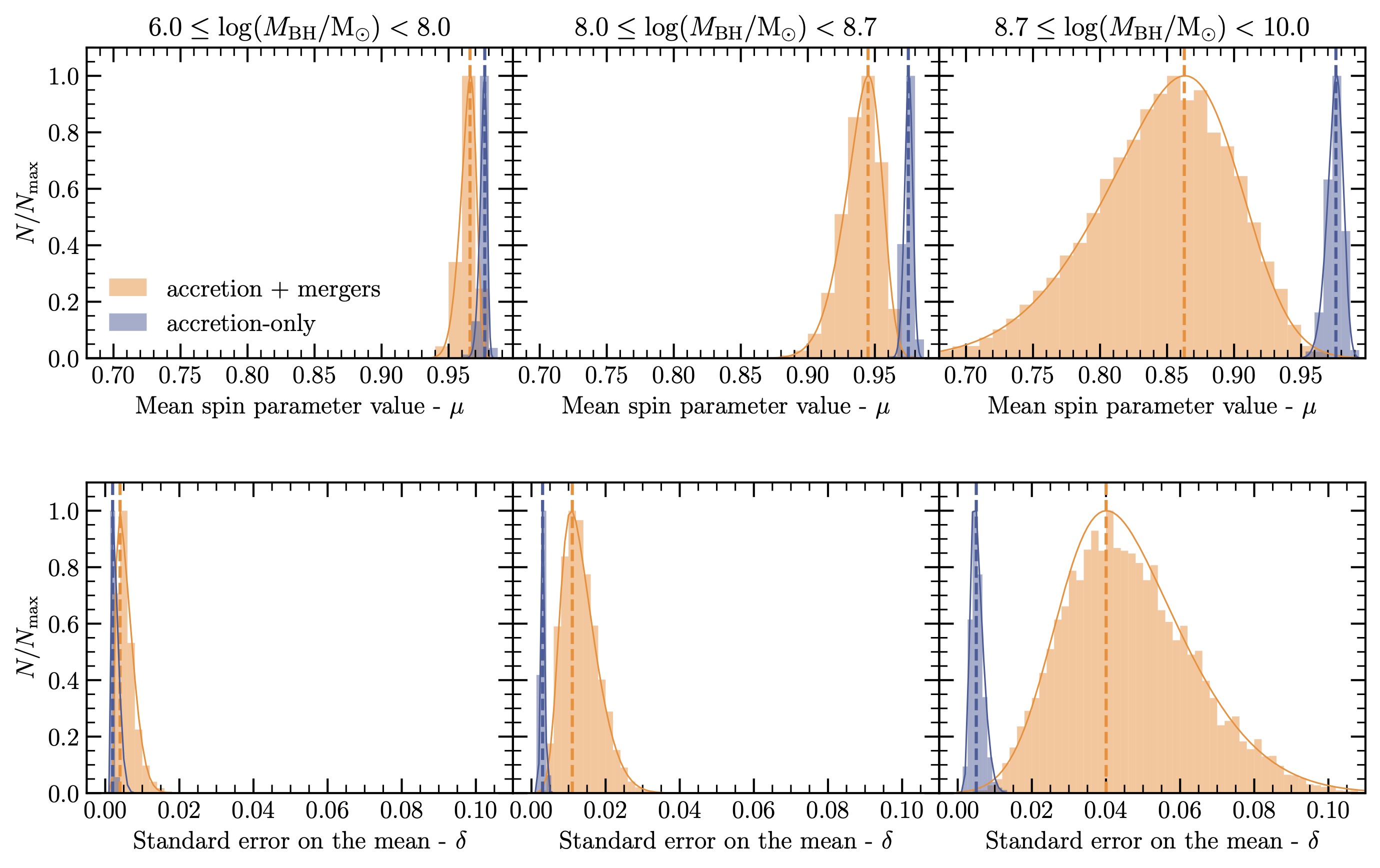}
    \caption{Summary of mean spin parameter estimates for $10^4$ realisations 
    of simulated \textit{HEX-P} observing campaigns. \textit{Top panels:}
    distribution of expected mean \spin\ estimates ($\mu$) in bins of \mbh\ for \acc\ (blue) and \accmer\ (orange) SMBH growth scenarios.
    \textit{Bottom panels:} corresponding distributions of the standard errors on the
    mean ($\delta$). Peak locations in each distribution
    are marked with vertical dashed lines. Differences between the mean spin
    measurements for the two scenarios become apparent for $M_{\rm BH} >
    10^8\, \rm{M_\odot}$ with the most massive bin showing the strongest signal.
    }
    \label{fig:mean-dist}
\end{figure}

Moving on from an individual realisation, the top panels in 
Fig.~\ref{fig:mean-dist} present the
distribution of mean \spin\ measurements ($\mu$) for the whole suite of $10^4$
simulations for the \accmer\ (orange) and \acc\ (blue) SMBH growth histories for
the three bins of \mbh. Vertical dashed lines mark $\mu$ at which each
distribution peaks, indicating the most likely expected mean spin parameter
measurement for each growth scenario. Similarly, bottom panels present the
corresponding distributions of the standard errors on the mean ($\delta$)
together with their peaks marked with vertical dashed lines. The mean spin
parameter values from Horizon-AGN together with their most likely measurements
expected from \hexp\ observations are summarised in Table~\ref{tab:mean-stats}.

Together Fig.~\ref{fig:mean-dist} and Table~\ref{tab:mean-stats} demonstrate
the potential of mean \spin\ measurements in bins of \mbh\ for differentiating
between the \acc\ and \accmer\ SMBH growth scenarios within the Horizon-AGN
cosmological model. As expected from Fig.~\ref{fig:tongues}, the simulated
constraints in the lowest mass bin are consistent between the two SMBH
populations, but become statistically separable for $M_{\rm BH} > 10^8 \rm
M_\odot$. In all three panels in Fig~\ref{fig:mean-dist}, $\mu_{\rm acc}$
distributions are narrower than in the \accmer\ case, with consistently
high peak values matching the true \radeff\ bias-corrected mean \spin\ in
Horizon-AGN to within $<2$\% . The orange distributions, on the other hand, 
peak at progressively lower $\mu_{\rm acc+mer}$ values with increasing \mbh, 
following the trend expected from the cosmological simulation to within
$<2$\% as well. We also note that the simulated \accmer\ mean spin measurements form
a~visibly left-skewed distribution primarily owing to the progressive spin parameter
uncertainty  prescription in our simulations, which assumes larger \dspin\
for lower spin values based on our \textit{HEX-P} spectral simulations
(Section \ref{sec:spectral-sims})\footnote{The skewness is also present in
the \acc\ scenario, however this effect is more not so visibly apparent due
to the narrow width of the distribution.}. The middle mass bin shows a
non-neglible overlap between the $\mu_{\rm acc}$ and $\mu_{\rm acc+mer}$
distributions, however their two peaks are clearly separated by $\Delta_\mu
\simeq 0.03$. For \mbox{$\log(M_{\rm BH}/\rm M_\odot) > 8.7$} the two SMBH
growth scenarios form virtually non-overlapping distributions with peak
$\mu$ separated by $\Delta_\mu \simeq 0.16$, despite the modest number of 9
BASS sources available in the mass bin. We note, however, that small sample statistics do
lead to a~significant spread in the simulated $\mu_{\rm acc+mer}$ values.

\begin{figure}[h!]
    \centering
    \includegraphics[width=0.5\textwidth]{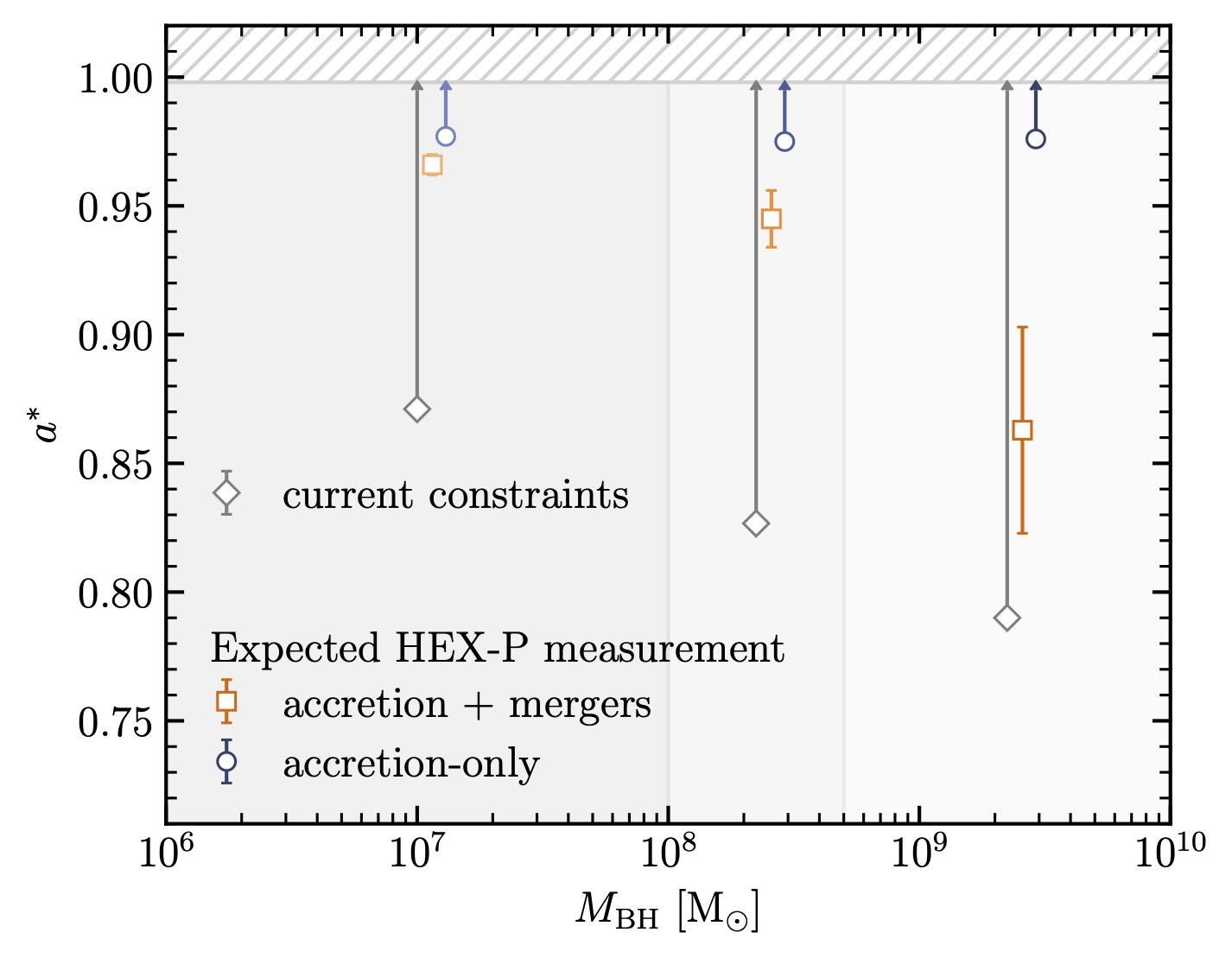}
    \caption{Comparison of mean spin parameter measurements in bins of \mbh\
    between published constraints (grey diamonds) and expected HEX-P
    measurements for \acc\ (blue circles) and \accmer\ 
    (orange squares) SMBH growth scenarios. For HEX-P measurements we show
    the most likely result from our simulations, corresponding to the vertical dashed
    lines in Fig.~\ref{fig:mean-dist}. Background shading marks our \mbh\ binning scheme, while the grey hatching at the top indicates spin parameter range forbidden by \cite{Thorne74} limit.  The expected \textit{HEX-P spin 
    parameter measurements will allow us to differentiate between SMBH growth 
    histories}, unlike \spinmass\ constraints placed until now.}
    \label{fig:main-result}
\end{figure}

In Fig.~\ref{fig:main-result} we summarise the prospects for constraining the 
cosmic growth history of supermassive black holes via X-ray reflection
spectroscopy with a \hexp\ spin survey similar to the design outlined above.
The figure shows the expected mean \spin\ measurements in three bins of \mbh\
for \acc\ (blue circles) and \accmer\ (orange squares) scenarios, inferred
from our suite of simulated observations of the brightest 100 AGN in our
parent BASS sample. The $y$-axis locus of individual points correspond to the vertical
dashed lines in the top panels of Fig.~\ref{fig:mean-dist}, while the length
of the error bars correspond to the vertical dashed lines in the bottom panels.
Since it is likely that observations of high-\spin\ SMBHs will result in
a significant fraction of constraints that are lower limits, we mark the
prediction for the \acc\ growth scenario as a lower limit. We also compare the
results of our simulations to the current constraints, showing mean
spin parameter values in identical mass bins for \spinmass\ measurements 
calculated earlier in Section~\ref{sec:obs-vs-sims} (and presented
earlier in Fig~\ref{fig:tongues}).

Fig.~\ref{fig:main-result} clearly demonstrates that \textit{future \hexp\
spin measurements will allow us to differentiate between the two SMBH growth 
channels as inferred from the Horizon-AGN cosmological model}. The expected
mean measurements in the highest mass bin alone reject the null hypothesis 
of the \acc\ and \accmer\ results being drawn from the same distribution
at the level of $2.8\sigma$ \mbox{($p_{[8.7-10.0]} = 2.5\times 10^{-3}$)}.
Across the whole suite of our simulations, $>82$\% ($>36$\%) of paired
simulated measurements in the highest mass bin reject the null hypothesis at
$>2\sigma$\ ($>3\sigma$) confidence level. When we combine the measurements for
both mass bins above $10^8 \rm M_\odot$ in SMBH mass, the expected \mbox{\hexp} measurement can differentiate between the two growth histories at $\sim 4.3\sigma$ and across all simulated measurements
at $>3\sigma$ ($>4\sigma$) confidence level in $>88$\% ($>65$\%) of all paired draws.
Fig.~\ref{fig:main-result} also illustrates that the current spin constraints
are not yet sufficient to make the distinction between these different growth
scenarios; a dedicated SMBH spin survey with improved individual constraints,
a larger sample size and well-controlled biases -- similar to the one
envisioned here for \textit{HEX-P} -- is really needed to drive the combined
SMBH spin constraints to the level that they are able to do so.

\section{Final Remarks} 
\label{sec:conclusions}

We have now confirmed that a~sample of 100 AGN with the data~quality determined in
Section \ref{fig:simspec} and the mass distribution of the 100 brightest sources
in our BASS selection (Section \ref{sec:bass-parent}) should be sufficient to distinguish
between different cosmological SMBH growth scenarios. Moving on, we finish our analysis with
an assessment of the level of observational investment from \textit{HEX-P} that
would be required to achieve this result. Given that a~150\,ks \textit{HEX-P}
exposure provides the necessary data~quality for our required spin constraints
($\sigma_{a^*} \leq 0.15$ for $a^* = 0.5$) for a~source with a~2--10\,keV flux
of $6 \times 10^{-12}$\,erg~cm$^{-2}$~s$^{-1}$, we calculate a~rough estimate for
how long it would take for \textit{HEX-P} to provide this level of data~quality
for all of these 100 sources by scaling this 150\,ks exposure to the actual
observed 2-10\,keV flux for each source (also reported in the BASS catalogue).
Formally this is a~slightly conservative approach, as the instrumental background
will make a~smaller contribution for higher source fluxes. Even then, for the
faintest of these sources the background already only impacts the highest
energies probed by each detector (see Figure \ref{fig:simspec}), so the
\textit{HEX-P} background is likely low enough to only have a~minimal
effect across this sample. Summing the exposure required for each of the
brightest 100 sources, we find that a~total observing investment of $\sim$\totexp\,Ms
from \textit{HEX-P} would be necessary.

While this is undoubtedly a~very significant investment, it is nevertheless a
program that would be achievable when spread over the full 5-year lifetime of
\textit{HEX-P}. In particular, a~spin survey such as this could serve a~similar
function to the \textit{HEX-P} mission as the ongoing \swift/BAT survey
conducted by \nustar\ (e.g. \citealt{Balokovic20}). This program gradually
builds up \nustar\ observations of hard X-ray sources detected by the
\textit{Swift}/BAT detector via~regular scheduling of short `filler'
observations, which are easily around the more extensive Guest Observer program.
Indeed, even the longest exposures required here -- 150\,ks -- could be split
into $3\times50$\,ks observations for ease of scheduling. The regular inclusion
of such observations in the schedule would also allow for increased flexibility
regarding the ability of \textit{HEX-P} to respond to transient phenomena~(i.e.
target-of-opportunity observations), as these survey observations can easily be
rescheduled, since they have no real time constraints and the simultaneity of
the coverage across the full 0.3--80\,keV bandpass is already guaranteed.

Most importantly, our study demonstrates that in the light
of theoretical predictions delivered by state-of-the-art
cosmological hydrodynamical simulations tracing SMBH 
spin evolution, only large statistical samples of consistent 
measurements are capable of isolating SMBH growth channels.
Once radiative efficiency-spin bias is accounted for, the 
differences in statistical properties of SMBH populations in the \spinmass\
plane decrease in magnitude, requiring both high precision on individual measurements 
and numerous AGN observations at $M_{\rm BH} > 10^8 \rm M_\odot$ to constrain
average trends in spin parameter as a~function of SMBH mass. Although
such measurements are, in principle, possible with currently available 
instruments, they critically require simultaneous observations with soft 
and hard X-ray observatories (e.g. \xmm\ and \nustar).  The necessary
exposure time comparable to the expected \hexp\ investment for 
these observatories renders such a study simply unfeasible, due to the 
limitations associated with program allocation and scheduling conflicts
among individual instruments. 

Beyond distinguishing cosmological BH growth scenarios via~the spin measurements,
the high-S/N broadband X-ray data~generated by such a~survey would represent an
undoubted legacy for the \mbox{\hexp} mission, and broader studies of AGN in
general. For example, inner disk inclinations will be well constrained (see Fig.
\ref{fig:simspec}), providing further tests of the broad applicability of the
Unified Model (\citealt{Antonucci1993}) as well as allowing for additional sanity
checks of the reflection results/searches for disk warps via~comparison of these
inclinations with constraints from the outer disk from e.g. VLT/GRAVITY (see the
cases of NGC3783 and IRAS\,09149-2461, where the inner and outer disk
inclinations from reflection studies and GRAVITY are in excellent agreement:
\citealt{Brenneman11, Walton20iras, Grav20iras09149, Grav21ngc3783}). These
observations would also provide good constraints on the X-ray corona~for a~large
sample of AGN, both in terms of its size scale (see Figure \ref{fig:simspec})
and its plasma~physics via~temperature measurements; all of these observations
will allow for good constraints on the electron temperature, facilitating further
tests of coronal models (e.g. \citealt{Fabian15}; see also \textbf{[Kammoun et al.
2023]}. All-in-all, while a~program of this nature would be a~major investment,
the scientific return provided would be suitably vast.

\section{Appendix}

\subsection{Radiative efficiency -- spin bias in flux limited samples}

A strong, positive, nonlinear relationship between the black hole spin parameter 
and the radiative efficiency in thin relativistic disks \citep{NovikovThorne1973} 
is frequently discussed in the literature as a potential source of observational 
bias for SMBH spin measurements, preferentially selecting sources with higher $a^\ast$. 
In order to account for this bias in our predictions from the Horizon-AGN
cosmological simulation, we consider the effect this $\eta(a^\ast)-a^\ast$ relation
has on a sample of X-ray measurements defined by by a minimum observable flux (i.e. a flux-limited sample) .
Following the dicussion in \cite{Brenneman11}, we assume that potentially observable
AGN are uniformly distributed within a~Cartesian space (an assumption appropriate
for the low-redshift Universe probed by our parent sample of BASS sources in 
Section~\ref{sec:bass-parent}), their spin parameter \spin\ is independent
of the instantaneous accretion rate \bhmdot\ and the X-ray spectra do not 
vary dramatically as~a~function of either \bhmdot\ or \spin.

\begin{figure}
    \centering
    \includegraphics[width=\textwidth]{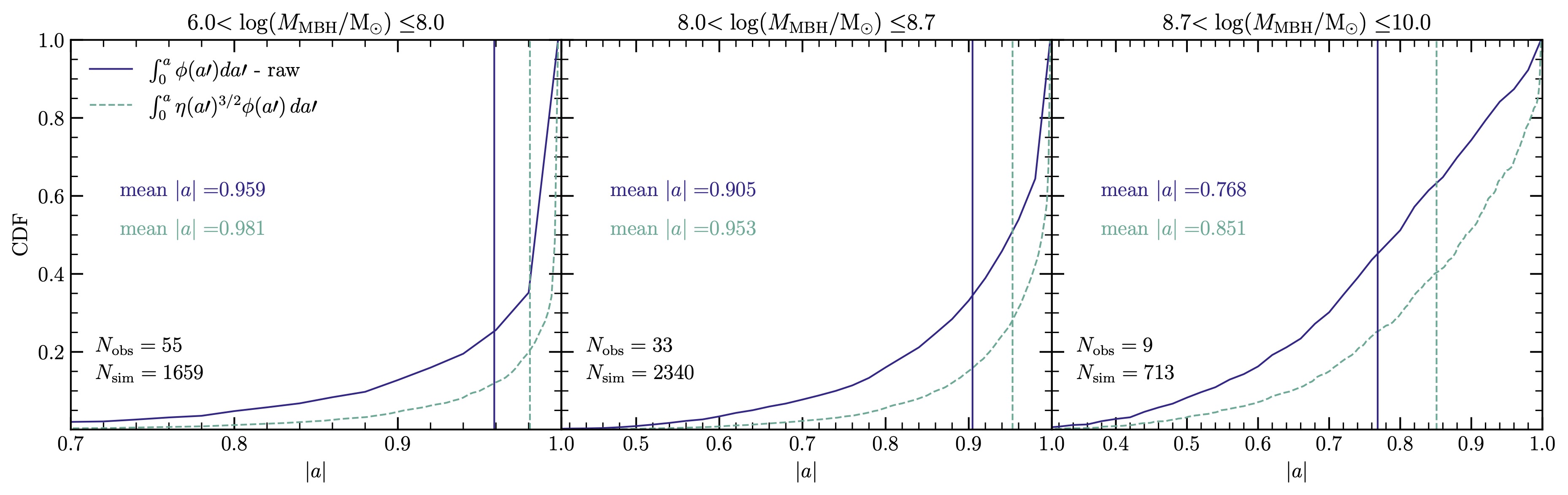}
    \caption{Change in the inferred cumulative distribution function (CDF) 
    and associated mean $a^\ast$ of the \accmer\ 
    Horizon-AGN SMBH population in response to the $\eta - a^\ast$ bias. 
    Solid lines show the raw CDF in each \mbh\ bin, while the dashed ones
    indicate the CDF resulting from $\eta-a^\ast$ correction. Vertical solid
    and dashed lines indicate inferred population means for the raw and 
    $\eta$-bias corrected distributions, respectively.
    Once the relationship between radiative efficiency and spin is accounted for, 
    inferred spin distributions are shifted towards higher $a^\ast$ values with
    the change most apparent in the highest mass bin.}
    \label{fig:horizonagn-lumbias}
\end{figure}

Under these assumptions, for a~given luminosity $L= \eta(a^\ast)\dot{M}c^2$, 
an X-ray flux limit effectively corresponds to a~limit in distance 
in a~Cartesian space ($r_L$), i.e:
\begin{equation}
    r_L = \qty(\frac{L}{4 \pi F})^{\flatfrac{1}{2}} 
    \propto \qty(\frac{L}{F_X})^{\flatfrac{1}{2}}\,.
    \label{eq:r_L}
\end{equation}
The total number of AGN in a given range of luminosity $L + dL$ and spin 
parameter $a^\ast~+ da^\ast$ is then given by:
\begin{equation}
    dN = V_L \Phi(L, a^\ast)dLda^\ast, \quad \text{where} \quad V_L=\frac{4}{3} \pi r_L^3
    \label{eq:dN_raw}
\end{equation}
and $\Phi(L, a^\ast) dLda^\ast$ is the spatial density of objects in the luminosity and 
spin increments. Combining Eq.~\ref{eq:r_L} and \ref{eq:dN_raw}, the number of AGN
in an incremental range in \spin\ and $L$ then has the following dependence on luminosity
in a flux-limited sample:
\begin{equation}
    dN \propto L^{\flatfrac{3}{2}} \Phi(L,a^\ast) dL da^\ast
\end{equation}
which from $L= \eta(a^\ast)\dot{M}c^2$ has the following dependency on $\eta(a^\ast)$ and \bhmdot:
\begin{equation}
    dN \propto \eta(a^\ast)^{\flatfrac{3}{2}} \dot{M}^{\flatfrac{3}{2}} 
    \Phi(\dot{M}, a^\ast) d\dot{M} da^\ast\, .
    \label{eq:dN_mdota}
\end{equation}
Assuming that \bhmdot\ and \spin\ are not mutually dependent, 
one can separate $\Phi$ into a product of unary functions of 
\bhmdot~and~\spin, namely $\Phi(\dot{M}, a^\ast)=f(a^\ast)g(\dot{M})$.
This way, Eq.~\ref{eq:dN_mdota} becomes
\begin{equation}
    dN \propto \eta(a)^{\flatfrac{3}{2}} f(a) da~\, 
   \dot{M}^{\flatfrac{3}{2}} g(\dot{M}) d\dot{M}\, .
\end{equation}

Finally, once we assume the dependence of $dN$ on \bhmdot\ can be marginalised over, 
the total number of AGN expected per \spin\ increment, $dN(a^\ast)=\rm{PDF}(a^\ast)$, 
can be expressed as
\begin{equation}
    dN(a^\ast) \propto \eta(a)^{\flatfrac{3}{2}} f(a^\ast) da^\ast~
    \int_{\Omega_{\dot{M}}} \dot{M}^{\flatfrac{3}{2}} g(\dot{M}) d\dot{M}
\end{equation}
and hence the final effect of the black hole spin dependence of radiative efficiency on 
the the cumulative distribution function of the expected spin measurements,
$\rm{CDF}(a^\ast)$, becomes:
\begin{equation}
    \text{CDF}(a^\ast) = \int_0^{a^\ast} dN(a^{\ast \prime})da^{\ast \prime} 
    =  \int_0^{a^\ast} \eta(a^{\ast \prime})^{\flatfrac{3}{2}} f(a^{\ast \prime}) da^{\ast \prime}
\end{equation}

Fig.~\ref{fig:horizonagn-lumbias} presents the effect of this radiative efficiency-spin bias
on spin parameter distributions in the \accmer\ SMBH population, 
inferred from the Horizon-AGN cosmological simulation. All three panels compare
the raw cumulative distribution functions, $\rm{CDF}(a^\ast)$, in solid 
blue lines against their expected observed counterparts from flux-limited surveys in teal dashed lines.
The solid and dashed vertical lines correspond to their respective mean spin 
parameter values. The figure is separated into three panels, showing the expected 
CDF in the three SMBH mass bins introduced in Sec.~\ref{sec:sample-reqs},
together with black hole population sizes in each bin expected from 
the brightest 100 BASS AGN in our parent sample, $N_{\rm obs}$, compared
against the corresponding number of SMBH found in Horizon-AGN, $N_{\rm sim}$. 
Fig.~\ref{fig:horizonagn-lumbias} clearly demonstrates how accounting for the 
relationship between the radiative efficiency and spin shifts the 
expected spin parameter distributions towards higher \spin\ values,
biasing the expected mean \spin\ measurement upwards 
by up to $\Delta a^\ast=0.083$ in the most massive bin.

\section*{Conflict of Interest Statement}

The authors declare that the research was conducted in the absence of any commercial or financial relationships that could be construed as a~potential conflict of interest.



\section*{Funding}
J.M.P. and J.A.G. acknowledge support from NASA grants 80NSSC21K1567 and 
80NSSC22K1120. The work of D.S. was carried out at the Jet Propulsion Laboratory, California Institute of Technology, under a contract with NASA. C.R. acknowledges support from the Fondecyt Regular (grant 1230345) 
and ANID BASAL (project FB210003). L.M. acknowledges support from the CITA National Fellowship.

\section*{Acknowledgments}



\bibliographystyle{Frontiers-Harvard} 
\bibliography{main}

\end{document}